\documentclass[a4paper,11pt]{article}

\title{\bf Thin shells of dust in a compact universe}
\author{Flavio Mercati\footnote{flavio.mercati@gmail.com}
\vspace{12pt} \\
\it \small Dipartimento di Fisica, Universit\`a di Roma ``La Sapienza''.\\
\it \small P.le A. Moro 2, 00185 Roma, Italy;
\vspace{6pt}
\\
\it \small Perimeter Institute for Theoretical Physics,\\
\it \small 31 Caroline Street North,
Waterloo, ON, N2L 2Y5 Canada}
\date{\today}

\usepackage{amssymb,bbold}
\usepackage{amsmath}
\usepackage[usenames,dvipsnames]{color}
\usepackage{tensor}
\usepackage{bbm}
\usepackage{graphicx}
\usepackage{cancel}
\usepackage[left=1in,right=1in,top=0.85in,bottom=1in]{geometry}                  
\usepackage{setspace}
\usepackage{hyperref}
\usepackage{color}
\usepackage{wrapfig}
\usepackage{lipsum}
\usepackage{mathrsfs}  
\usepackage{floatrow}
\usepackage{bm}

\def\Xint#1{\mathchoice
   {\XXint\displaystyle\textstyle{#1}}%
   {\XXint\textstyle\scriptstyle{#1}}%
   {\XXint\scriptstyle\scriptscriptstyle{#1}}%
   {\XXint\scriptscriptstyle\scriptscriptstyle{#1}}%
   \!\int}
\def\XXint#1#2#3{{\setbox0=\hbox{$#1{#2#3}{\int}$}
     \vcenter{\hbox{$#2#3$}}\kern-.5\wd0}}

\def\dashint{\Xint-}

\hypersetup{
    colorlinks=true,         
    linkcolor=blue,          
    citecolor=red,        
    urlcolor=Violet             
}

\newcommand{\sfrac}{\textstyle \frac}
\newcommand{\st}[1]{\text{\tiny \rm #1}}

\renewcommand{\d}{{\rm d}}

\newcommand{\widfigs}{0.29}

\begin{document}

\maketitle

%

\begin{abstract}
I present the first analytical study of gravitational collapse in a compact CMC foliation with $S^3$ spatial topology. The solutions I find, in this context, will be both solutions of Shape Dynamics and General Relativity. The aim is to  describe a system undergoing gravitational collapse in Shape Dynamics, so a well-justified and useful  simplification is to assume spherical symmetry. This kills all the local gravitational degrees of freedom, but some nontrivial degrees of freedom are recovered by introducing matter. The simplest form of matter is infinitely thin spherical shells of dust, of which I need at least two in order to have a nontrivial dynamics. With a single shell the system is dynamically trivial, but it nevertheless admits a solution which represents a `frozen' shell at equilibrium in a globally de Sitter universe. Such a solution is, to my knowledge, new.  I am able to solve analytically also the case with two shells, which has a nontrivial dynamics. When the rest mass of one shell is much smaller than the other, the system is suitable to model a  compact universe in which one subsystem (the `light' shell) undergoes gravitational collapse while the rest of the matter (the `heavy' shell) plays the role of spectator. It turns out that, if the cosmological constant is zero or positive but small, and the rest mass of the two shells are sufficiently different, when the `light' shells collapses the ADM  equations become ill-defined and cease to admit a solution. The shape-dynamical description, however, seems still well defined and can be continued past this point, possibly signalling a departure of Shape Dynamics from exact equivalence with General Relativity.
\end{abstract}

\newpage

\section{Introduction}

Shape Dynamics (SD) describes gravity as the dynamics of 3D conformal geometry. Fundamentally, the theory does not involve the concept of spacetime, and has no inkling of  relativity of simultaneity.  Such concepts should emerge effectively, on-shell~\cite{gryb:shape_dyn,Gomes:linking_paper,FlaviosSDtutorial}, while conformal invariance is implemented off-shell as well. In other words, the fact that there is no preferred hypersurface of simultaneity (if true) can at most be a property of the solutions of the theory;\footnote{ In fact, it is already the case that the ADM Hamiltonian, or scalar constraint, corresponds to re-definitions of surfaces of simultaneity only for space-times that  satisfy the Einstein equations~\cite{Lee_Wald_OffShell_refoliation}.}  it is a dynamical property, as opposed to  conformal invariance, which is a kinematical property. 

Abdicating the spacetime view, Shape Dynamics requires  different tools to define the dynamics. In principle, any conformally-invariant global Hamiltonian would suffice, but in general such Hamiltonians won't be phenomenologically viable. Among the possible choices, one has the special property that it generates a dynamics which, in a certain conformal gauge (determined by the `Lichnerowicz--York equation'), is equivalent to that of General Relativity (GR), and therefore inherits an extensive endowment of experimental support from Einstein's theory. This Hamiltonian is the by-product of York's conformal method for solving the initial value problem of GR~\cite{York1971,FlaviosSDtutorial}. Ideally, we would like to keep agnostic about the Hamiltonian and study the whole space of conformally- and diffeomorphism-invariant Hamiltonians. If the existence of universality classes and finite-dimensional critical surfaces with finite fixed points could be proven, then one would have a consistent quantum theory of conformal geometrodynamics. The hope is that, if such a program is successful, the point in theory space that corresponds to York's Hamiltonian lies on such a critical surface, and therefore belongs to an asymptotically-safe orbit of the RG flow. Then our quantum theory of Shape Dynamics would also admit the correct classical limit.

Until we reach the level of technical development that is necessary to put to the test the above conjecture, it is important to study the classical theory of gravity that is defined by the York Hamiltonian. The solutions of this theory are equivalent to solutions of Einstein's equations only when the latter describe a spacetime that can be foliated by constant-mean-extrinsic curvature (CMC) slices. This condition is pretty generic and is satisfied in most usual physical situations (and, in particular, in all situation we have direct experimental access to). However there are space-times which satisfy Einstein's equations and admit no such complete foliation. In such situations the conformally-invariant dynamics defined by the York Hamiltonian may still be well-defined. For instance, by assuming continuity of the shape degrees of freedom, it was shown in \cite{ThroughTheBigBang} that Shape Dynamics resolves the Big Bang singularity, and allows to evolve smoothly through it. This comes at the cost of renouncing, already at the classical level, to the requirement of a smooth spacetime. Instead one has two spacetimes which are `soldered' at a  singular hypersurface.

It is natural to ask whether SD can similarly resolve other kinds of GR singularities. The first case we should study with this aim is that  of the Schwarzschild singularity. This issue was investigated ia series of papers:
\begin{enumerate}
\item[\textbf{1.}]  In~\cite{Birkhoff_SD} H. Gomes studied the vacuum, asymptotically flat, spherically symmetric solutions of SD. The result was an odd-lapse  maximal slicing of Schwarzschild spacetime, which covers the first and third quadrants of the Kruskal extension, and therefore avoids the two quadrants which contain singularities. The spatial slices look like a wormhole metric: they have two asymptotically flat ends, and  in between a `throat', that is, a minimal-area sphere.
This initial result had the following shortcomings:
\begin{enumerate}
\item[\textbf{(a)}] The spherically-symmetric ansatz implies that the spatial geometry is conformally flat. Therefore, by definition,  there can be no local gravitational shape\footnote{\emph{I.e.} conformally-invariant.} degrees of freedom. In absence of matter, and with the boundary conditions fixed by hand, this is not a genuine  shape-dynamical system.
Moreover, it is not clear that the result of~\cite{Birkhoff_SD}  should really represent a black hole in Shape Dynamics. The physically relevant question is whether such an object would form as the result of the gravitational collapse of ordinary matter.

\item[\textbf{(b)}]  The boundary conditions chosen at infinity are arbitrary. They require the following falloff conditions for the metric and its conjugate momenta: $g_{ij} = \delta_{ij} + \mathcal{O} (r^{-1}) $, $p^{ij} = \mathcal{O} (r^{-2}) $. These are standard in GR literature~\cite{beig1987poincare}, but they are not natural in SD; they amount to requirements on the asymptotic structure of spacetime (not of the conformal geometry).
Only spatially compact solutions are truly relational,\footnote{For Einstein~\cite{EinsteinNote}, in the spatially compact case ``the series of causes of mechanical phenomena is closed''.} and asymptotic flatness can be at best an approximation of an empty region inside a compact universe~\cite{Tim_Proceedings_TheoryCanada9}.

\item[\textbf{(c)}]  In \cite{Birkhoff_SD}  \emph{maximal slicing,} $g_{ij} p^{ij} = 0$ was used, instead of the constant-mean-extrinsic curvature condition $g_{ij} p^{ij} = \langle p \rangle \sqrt g$ (where $ \langle p \rangle = \int d^3 x g_{ij} p^{ij}  / \int d^3 x \sqrt g $ is a spatial constant). This is justified for asymptotic flatness, seen as an approximation to a  small empty region in a much larger universe. In fact, in the relevant equations (see Eqs.~(\ref{SphericalADMconsts}) and (\ref{SolConstraints}) below), the terms that depend on $ \langle p \rangle $ (as well as those that depend on the cosmological constant $\Lambda$) go like the sixth power of the areal radius of the metric, $\sqrt{g_{\theta\theta}}$. All the other terms depend on lower powers of  $\sqrt{g_{\theta\theta}}$, and therefore  dominate near the origin  $\sqrt{g_{\theta\theta}} =0 $. From this point of view, discarding any dependence on $ \langle p \rangle $ seems justified. However, it also means that the asymptotically flat solution of~\cite{Birkhoff_SD} corresponds to an infinitely thin slice of CMC time $ \langle p \rangle $ (also called York time).
\end{enumerate}

\item[\textbf{2.}]  The result of~\cite{Birkhoff_SD} generated a number of  spinoffs (\emph{e.g.}~\cite{Gabe_Kerr,Gabe_Parity_Horizons, ThinshellPaper}).  In~\cite{ThinshellPaper}, together with H. Gomes, T. Koslowski and A. Napoletano, I  partly addressed  issue \textbf{(a)}, by studying  the simplest form of spherically-symmetric matter: an infinitely thin spherical shell of dust, while keeping all the other assumptions of~\cite{Birkhoff_SD} (standard GR asymptotically flat fall-off conditions,  maximal slicing, no cosmological constant). With this setup the system has one pair of Hamiltonian degrees of freedom (the radius of the shell and its radial momentum). The system with a single shell still has no relational matter degrees of freedom, as there isn\rq{}t a second matter subsystem to stand for comparison. It is therefore not a genuine shape-dynamical system. Nonetheless, one could interpret this system as a background over which weak perturbations can propagate~\cite{Tim_Proceedings_TheoryCanada9}. These perturbations can give rise to an arbitrary number of genuine degrees of freedom,  `probing' the background without influencing it too much. The solution we found, where it exists in phase space,  can therefore be a  good approximation to \emph{bona fide} shape-dynamical solutions, and it thus makes sense to draw (limited) physical conclusions from it.
In particular, in~\cite{ThinshellPaper} we showed that the `wormhole' geometry described in~\cite{Birkhoff_SD} is generated outside of the shell as a result of its collapse. The dynamical orbits of the shell in its reduced phase space were found, and could be classified as closed or open (depending on whether the shell has enough kinetic energy to reach escape velocity), as expected. In the same paper it was also found that the  radius of the shell reaches the throat only at the boundary of phase space - when the momentum diverges - requiring an infinitely long phase-space curve to do so. This implies that, in the no-backreaction approximation, the matter outside the shell goes through an infinite amount of change before the shell crosses the throat.

\item[\textbf{3.}]  In view of problem \textbf{(b)},  in~\cite{BirkhoffFlavio} I critically reassessed the boundary conditions assumed in~\cite{Birkhoff_SD}. Their use in GR is justified by the requirement of Poincar\'e invariance of the falloff conditions~\cite{beig1987poincare}, but in SD one is only authorized to assume symmetries  of the spatial slices at infinity, not of the spacetime metric. A closer look at the problem revealed the presence of an integration constant of the spherically-symmetric equations of SD. This parameter, called $A$ (sometimes called \emph{Estabrook--Wahlquist time}~\cite{Alcubierre}), is a spatial constant but can be time dependent, and it is set to zero by the boundary conditions assumed in~\cite{Birkhoff_SD}. It turns out that, if $A\neq 0$, the falloff conditions lose their invariance under asymptotic Lorentz transformations, but they are still invariant under spatial rotations and translations, and under time translations. Lorentz invariance of the boundary is not a legitimate request for an asymptotically flat solution of Shape Dynamics: it is only the spatial slices which have to develop the isometries of Euclidean space at the boundary.  So the parameter $A$ cannot be put to zero in the same way as in GR,  and has to be kept as an arbitrary function of time. However, the condition $A=0$  seems to be required in order to associate finite charges to spatial asymptotic rotations when strict spherical symmetry is relaxed.

This was pointed out in ~\cite{Henrique_Poincare_invariance}, where it was shown that the standard asymptotically flat falloff conditions of GR (which imply $A=0$) are necessary in order to ensure the well-posedness of the variational problem. In other words, if we relax the assumption of spherical symmetry, falloff conditions that allow $A \neq 0$ will attribute infinite values to some of the boundary charges (like angular momentum), which means that one cannot define counterterms that make the action differentiable. As soon as we depart from perfect spherical symmetry, asymptotically flat SD with $A\neq 0$ is not a well-defined dynamical system. This seems to be a powerful argument in favour of fixing $A=0$, however, as is shown in the present paper, in a closed universe this argument doesn't hold (there are no boundary charges and the variational problem is always well-posed), and the integration constant $A$ may admit values other than zero (it is determined by the state of motion of matter and setting it to zero `by hand' is inconsistent). Therefore, if asymptotically flat SD with $A\neq 0$ turns out to be inconsistent, it cannot be a good approximate description of a nearly-empty region in a larger closed universe. The issue of what is the right noncompact model of such a situation will be discussed in future works.

Although the boundary charges of non-spherically symmetric configurations can be infinite, in the perfectly spherically-symmetric case all the charges related to Euclidean symmetries (translations and rotations) are zero~\cite{BirkhoffFlavio}. Moreover, the charge associated to dilatations turns out to be proportional to the integration constant $A$. This suggests a physical meaning for $A$: if our asymptotically flat region is an approximation to an empty bubble in a larger universe, then the state of motion of the matter outside the bubble determines the boundary conditions. If the matter outside is expanding or contracting, this breaks the Lorentz invariance of the falloff conditions because it introduces a preferred frame. It does not, however,  break the translation- or rotation-invariance of the fall-off conditions. An asymptotically flat model cannot predict the value of $A$ at each instant, as it is the consequence of the dynamics of the rest of the matter in the universe (whether the bubble is expanding or collapsing), so we have to go beyond this approximation if we want to deal with a dynamically closed system. 

\end{enumerate}
In this paper I will finally face all of the shortcomings of the previous attempts, and attack a problem which is truly relational. This means that the spatial manifold will be compact, which addresses shortcoming \textbf{(b)}. The simplest topology we can choose is $S^3$, whose symmetry group $SO(4)$ can be broken into $SO(3)$ by the introduction of two antipodal poles, around which we assume rotational symmetry. This assumption of course deprives us of all local propagating degrees of freedom of the metric, exposing us to criticism \textbf{(a)}, which, due to constraints on mathematical tractability at the moment,  we can only address by introducing some form of matter. As in~\cite{ThinshellPaper}, the simplest choice is thin shells of dust. Considering only a single shell leads to a trivial solution, in which the shell has no dynamics (although it leads to a consistent solution of Einstein's equations which, to my knowledge, has not been found before - see Sec.~\ref{SubsecSingleShellUniverse}), so the minimum number of shells is two. Finally, by considering a compact universe, we have implicitly  introduced a finite volume for it, which is now itself a dynamical degree of freedom (at least in the ADM-in-CMC-gauge description of the problem), and for consistency we need to include its canonically conjugate degree of freedom: the York time $\langle p \rangle$. This addresses criticism \textbf{(c)}. As it turns out (see Sec.~\ref{SecDescriptionOfTheSolution}), in order to have a compact spatial manifold, we need to introduce also a cosmological constant, which however does not complicate the equations any further. 
In Sec.~\ref{SecDescriptionOfTheSolution} and \ref{ThinShellSection} I study this problem. I can solve  the system exactly, and describe analytically the on-shell surface representing reduced phase space for any possible value of the free parameters in the system. My analysis reveals a feature which will be the focus of the final Section~\ref{SecTheProblem}. This feature appears precisely in the cases which are the central focus of the present investigation: when the system is a good model of gravitational collapse, \emph{i.e.} when the two shells have very different rest masses, and the lightest one collapses to a small region in a  universe with a  small positive cosmological constant. In other words, when we expect to approach the formation of a black hole. In these cases I explicitly see a departure of the dynamics of SD from that of GR: as the collapse proceeds, at some point the description of the system in terms of ADM-in-CMC variables fails (there is no real solution of the Lichnerowicz--York equation, and consequently no CMC-foliated spacetime). This result is highly significant, as it indicates that Shape Dynamics might have more to say about black holes  than GR, just like it did in the case of the big-bang singularity in  the recent~\cite{ThroughTheBigBang}.

\section{Vacuum constraints and equations of motion}\label{SecDescriptionOfTheSolution}

\subsection{Spherically symmetric vacuum ADM constraints}

\begin{figure}[ht!]
\floatbox[{\capbeside\thisfloatsetup{capbesideposition={right,center},capbesidewidth=0.46\textwidth}}]{figure}[\FBwidth]
{\caption{\footnotesize On the right-hand side we plot a possible  shape for the polynomial  $\mathscr P$ versus $\sqrt{\sigma}/|m|$ on the vertical axis. In parallel, on the left, we plotted two possible choices of $\sigma$ as a function of $r$. The intervals in which $\mathscr P>0$ from the left-hand side plot are projected onto the vertical axis of this last diagram, so that one can see that $\sigma$ is confined within these intervals, and approaches their boundaries with zero derivative. Notice how in the upper interval, which is bounded from above and below, one can fit an arbitrary number of extrema of $\sigma(r)$, while the lower interval, which is only bounded from above, $\sigma$ can have only one extremum (a maximum) and otherwise has to go to zero.}\label{DomainOfSigmaFig}}
{\includegraphics[width=0.54\textwidth]{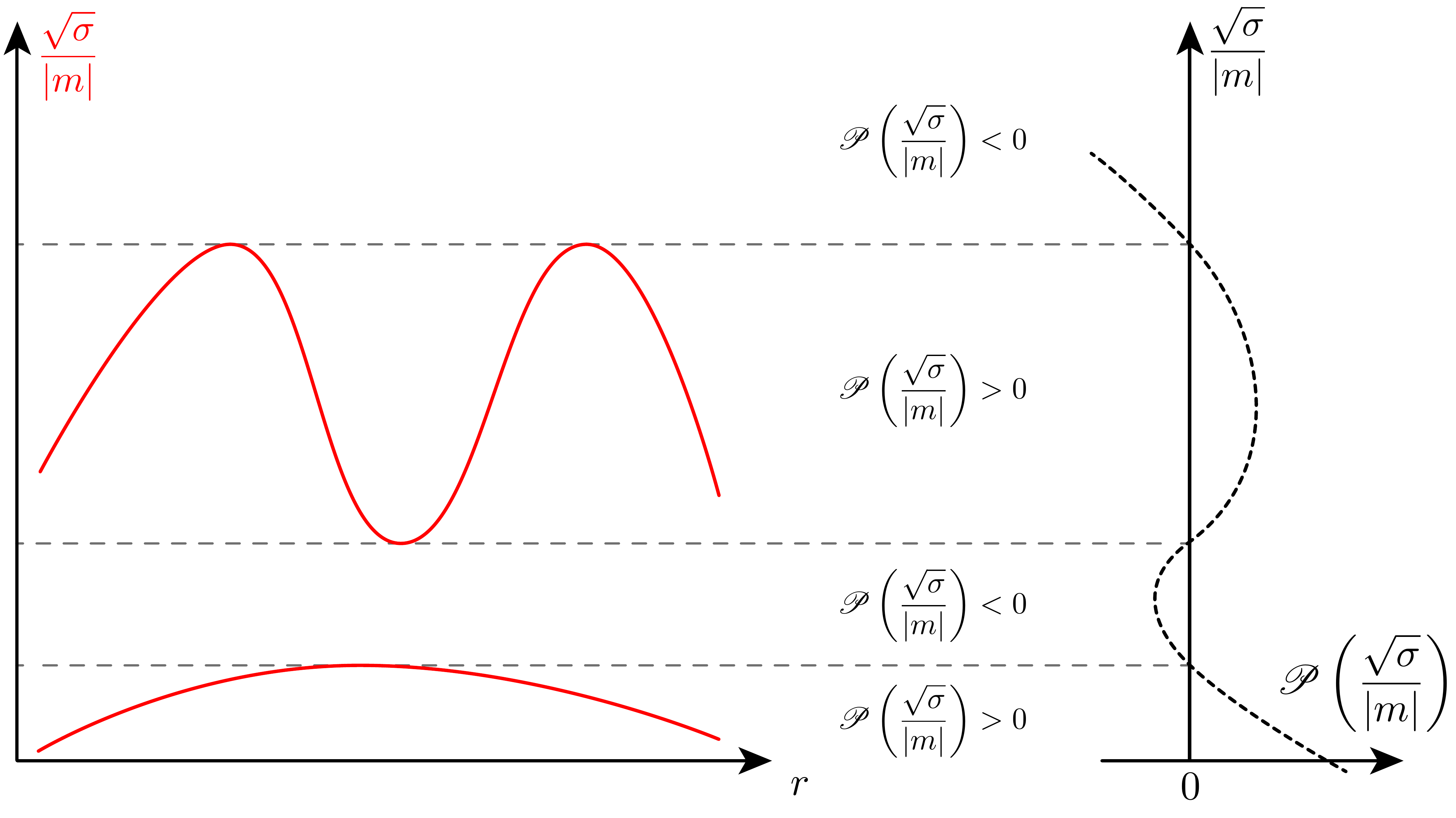}}
\end{figure}

Assuming a spherically-symmetric ansatz~\cite{ThinshellPaper,FlaviosSDtutorial} on the coordinate patch\footnote{In this paper $r \in [0,\pi]$, where $r=0$ and $r=\pi$ are the coordinates of, respectively, the south and north pole. It is customary to attribute $r$ an infinite range when modelling noncompact manifolds, and a finite range in the compact case. Of course, these are only unphysical coordinate choices as we may as well describe a compact manifold with a noncompact coordinate patch or vice-versa.} $r \in [0 , \pi]$, $\theta \in [0,\pi]$, $\phi \in [0,2\pi)$, the spatial metric $g_{ij}$ and its conjugate momentum $p^{ij}$ depend each on two functions of the radial coordinate $r$  only, respectively $\mu$, $\sigma$ and $f$, $s$:
\begin{equation}
g_{ij} = \text{diag} \, \left\{  \mu^2 , \sigma  , \sigma \, \sin^2 \theta  \right\} \,,
\qquad 
p^{ij} =  \text{diag} \, \left\{  \frac{f}{\mu} , {\frac s 2}     , {\frac s 2}   \, \sin^{-2} \theta  \right\} \, \sin \theta \,.
\end{equation}
The vacuum ADM Hamiltonian constraint $\mathcal H$ and  diffeomorphism constraint $\mathcal H_i$, and the volume-preserving conformal constraint $\mathcal C$ are~\cite{FlaviosSDtutorial,ADM}:
\begin{equation}\label{ADMconsts}
\mathcal H  =  \frac{1}{\sqrt g} \left( p^{ij} p_{ij} - {\frac 1 2} p^2 \right)
+ \sqrt g  (2\Lambda - R)   \,,
\qquad
\mathcal H_i  = -2 \, \nabla_j p^j{}_i  \,,
\qquad
\mathcal C = p - \langle p \rangle \, \sqrt g \,,
\end{equation}
after replacing the spherically-symmetric ansatz, they turn into
\begin{equation}\label{SphericalADMconsts}
\begin{aligned}
\mathcal H  &=  - \frac{1}{6 \sigma \mu ^2} \left[ \sigma ^2 \mu s^2 +  4 f^2 \mu ^3 -4 f \sigma \mu ^2 s + 12 \sigma \mu \sigma '' - 12 \sigma \sigma ' \mu ' - 3 \mu (\sigma')^2 
\right.
\\
& \left. \qquad \qquad ~~ - 12 \sigma \mu ^3 -( \langle p \rangle^2 - 12 \Lambda) \sigma ^2 \mu ^3  \right] \approx 0\,,
\\
\mathcal H_i &= \delta^r{}_i \left(  \mu f' - {\sfrac 1 2} s \sigma' \right) \approx 0 \,, 
\qquad  \mathcal C = \mu f + s \sigma  - \langle p \rangle \, \mu \, \sigma \approx 0 \,,
\end{aligned}
\end{equation}
where $'$ denotes the $r$-derivative.  These equations can be solved basically in the same way as we did in~\cite{BirkhoffFlavio} and~\cite{ThinshellPaper}. The explicit solution is:
\begin{equation}\label{SolConstraints}
 \begin{gathered}
 s  = \langle p \rangle \, \mu   - \frac{\mu}{\sigma} \, f  \,,
\qquad \qquad
 f = {\sfrac 1 3} \langle p \rangle \, \sigma  + \frac{A}{\sqrt \sigma}\,,\\
 \mu^2 =  \frac{(\sigma')^2}{\frac{A^2}{\sigma }  + \left( {\sfrac 2 3} \langle p \rangle  A - 8 \, m  \right) \sqrt{\sigma}  + 4 \, \sigma - {\sfrac 1 9}  \left( 12 \, \Lambda - \langle p \rangle^2 \right) \sigma^2 } \,.
\end{gathered}
 \end{equation}
The solution introduces two integration constants:\footnote{The constraint $\mathcal C \approx 0$ is algebraic and gives rise to no integration constants. Moreover, the equation $\mathcal H \approx 0$ is second-order while $\mathcal H_i \approx 0$ is first-order, so we should have a total of 3 integration constants. However, by finding a first integral of  $\mathcal H_i \approx 0$ and setting its value to $m$, we converted  $\mathcal H \approx 0$ into a first-order equation (the last of Eqs.~\ref{SolConstraints}), which admits one further integration constant~\cite{ThinshellPaper}.} $m$, which is the Misner--Sharp mass~\cite{BirkhoffFlavio,MisnerSharpMass}, and $A$, which is associated to the dilatational momentum of the boundaries of the empty region under consideration~\cite{BirkhoffFlavio}. The last of the three solutions relates $\mu$ to  $\sigma$ and $\sigma'$. Solving it requires choosing a radial diffeomorphism gauge. For example, in `isotropic' gauge $\sigma = \mu^2 \sin^2 r$ the relation coincides with the Lichnerowicz--York equation~\cite{FlaviosSDtutorial}, whose solution would introduce a further integration constant, which from now on we will call $k$. However this equation cannot be solved analytically, as it involves the inversion of the solution of an elliptic integral. Other radial gauge-fixings lead to a solvable equation, for example one can explicitly specify the form of the function $\sigma (r)$ (which is the square of the areal radius of the metric), which fixes automatically also the form of $\mu$. However it turns out that not every function $\sigma (r)$ is acceptable (one says that not all gauges are \emph{attainable}). This is because the last of Eq.~(\ref{SolConstraints}) is not compatible with every value of $\sigma$: the image of the function $\sigma(r)$ has to belong to the domain of positivity of the quantity $\left[\frac{A^2}{\sigma }  + \left( {\sfrac 2 3} \langle p \rangle  A - 8 \, m  \right) \sqrt{\sigma}  + 4 \, \sigma - {\sfrac 1 9}  \left( 12 \, \Lambda - \langle p \rangle^2 \right) \sigma^2\right]$, otherwise $\mu$  is imaginary and the metric ends up being Lorentzian. 
The quantity above is $\frac {4 m^4}{\sigma}$ times the following dimensionless polynomial:
\begin{equation}\label{MordorPoly}
\mathscr P [z] =  {\sfrac{1}{36}}  \left(6 C+\tau  z^3\right)^2 - (\pm 2\, z^3) - {\sfrac 1 3} \lambda  \, z^6  + z^4\,,
\end{equation}
where the sign $+$ corresponds to $m > 0$ and $-$ corresponds to $m < 0$. The quantities
\begin{equation}
z = \frac{\sqrt \sigma}{|m|} \,, ~~~ C = \frac{A}{2 \, m^2} \,, ~~~  \tau = |m| \, \langle p \rangle \,, ~~~ \lambda = m^2 \Lambda\,.
\end{equation}
are dimensionless.
So $\sigma$ has to be such that $\mathscr P [ \sqrt{\sigma/m^2}] >0$. Moreover, $\sigma$ can reach the border of this domain, where $\mathscr P [ \sqrt{\sigma/m^2}] =0$, but only in such a way that the quantity $\sigma (\sigma')^2 / \mathscr P [ \sqrt{\sigma/m^2}] $ (which is proportional to $\mu^2$) is positive, which means that $\sigma'$ has to be zero at the border. $\sigma'$ cannot be zero anywhere else, because that would make $\mu$ zero too and the metric would be degenerate, so we conclude that $\sigma$ has to be monotonic in the bulk of the domain of positivity of $\mathscr P$, and can have extrema only at the boundary of that region.
If we were able to analytically solve the isotropic LY equation, the same conditions would be satisfied automatically. In Fig.~\ref{DomainOfSigmaFig} we show what the form of an acceptable choice of $\sigma$ must be.

The $\mathscr P$ polynomial depends on several parameters: the areal radius $\sqrt{\sigma}$, the integration constants $A$ and $m$, the cosmological constant $\Lambda$ and York time $\langle p \rangle$. 
In Appendix~\ref{MordorAppendix} we study the region of positivity of the polynomial $\mathscr P$ in full generality, for any value of these parameters. The parameter $|m|$ can be used as a scale to make all the other parameters dimensionless. The only two parameters that are suitable for that role are $|m|$ and $|\Lambda|$ because, as we show below, they are both conserved. Choosing $|m|$ as the scale means that one has one time-independent dimensionless parameter $\lambda = m^2 \Lambda$ that should be fixed and gives different profiles for the `forbidden' region of $\mathscr P <0$. The other three dimensionless parameters, $z = \frac{\sqrt \sigma}{|m|}$,  $C = \frac{A}{2 \, m^2}$ and $ \tau = |m| \, \langle p \rangle$ are dynamical and take all possible values.
In Appendix~\ref{MordorAppendix} below we study the region $\mathscr P <0$ in the 3D space $ \left( z , C ,\tau\right)$, for any possible choice of value of $\lambda =  m^2 \Lambda$. Here is a particular example:
\begin{figure}[ht!]
\includegraphics[width=0.35\textwidth]{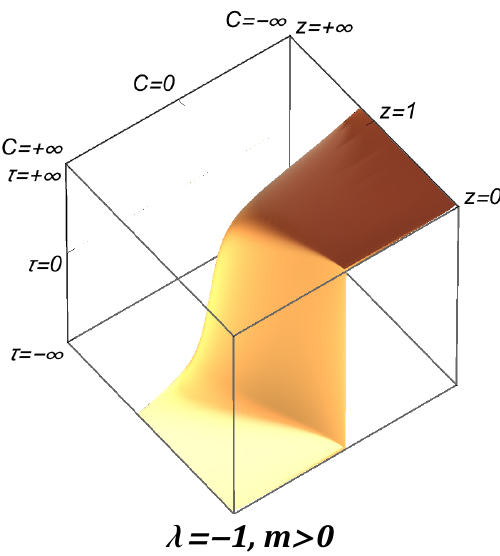}~~~~~~~~~~~~~~~~\includegraphics[width=0.35\textwidth]{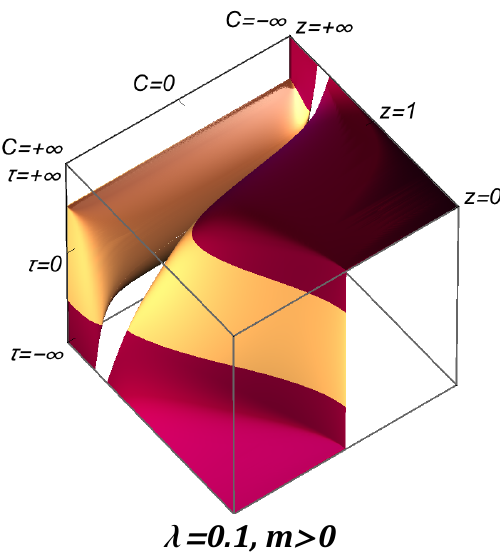}
\caption{\footnotesize The `forbidden' region $\mathscr P (z) < 0$ for \textbf{positive Misner--Sharp mass}  and  \textbf{negative}  (left) or \textbf{positive} (right) \textbf{cosmological constant.}  The part of the surface where $\tau^2 < 12 \Lambda$ is in yellow, while $\tau^2 > 12 \Lambda$ is in red.
}\label{MordorFig2_example}
\end{figure}

\subsection{Equations of Motion}

The ADM equations of motion require previous calculation of the CMC lapse, which is given by the Poisson bracket between $\mathcal H$ and $\mathcal C$,
\begin{equation}\label{GeneralLFE}
\left( 8  \Delta -2 R +12 \Lambda  - \langle p\rangle ^2 \right) N 
- {\sfrac 6 g} \left(p^{ij} -{ \sfrac 1 3} g^{ij}  \, p \right)\left(p_{ij} -{ \sfrac 1 3} g_{ij}  \, p \right) N
= \left\langle \sqrt g ~ \text{\it lhs} \right\rangle \,,
\end{equation}
where $ \left\langle \sqrt g ~ h \right\rangle = \frac{\int \sqrt g \, h(x) \,d^3x}{\int \sqrt g \, d^3x} $ is the spatial average and lhs stands in for the content of the left hand side of the equality, repeated under the mean sign. 
Under the assumption of spherical symmetry  [which for a scalar function like the lapse is just $N=N(r)$], Eq.~(\ref{GeneralLFE}) reduces to
\begin{equation}\label{SphSymmLFE1}
\begin{aligned}
\left(\frac{4 f     s }{\mu   \sigma  }  -\frac{4 f ^2   }{\sigma  ^2} -\frac{4    \mu'  \sigma' }{\mu^3 \sigma  }+\frac{4    \sigma'' }{\mu  ^2 \sigma  }-\frac{   (\sigma')^2}{\mu^2 \sigma^2}-\frac{4   }{\sigma}-\frac{   s^2}{\mu^2} \right)  N  +
\\
\left(12 \Lambda  - \langle p \rangle^2 \right) N  - \left( \frac{8 \mu'}{\mu  ^3}+\frac{8 \sigma' }{\mu^2 \sigma  }\right) N' + \frac{8 N'' }{\mu^2}
 =\left\langle \sqrt g ~ \text{\it lhs} \right\rangle  \,.
\end{aligned}
\end{equation}
The above equation can be formally solved by
\begin{equation}\label{SolutionLFEtwinshell}
N = \frac{\sigma'}{2 \mu \sqrt{\sigma}} \left( c_1 + c_2 \dashint \frac{\mu^3}{(\sigma')^2} \d r  + \frac w 6 \dashint \frac{\sigma^{3/2} \mu^3}{(\sigma')^2} \d r \right) \,,
\end{equation}
where $\dashint$ is the principal-value integral (see~\cite{ThinshellPaper,FlaviosSDtutorial} for the reason behind the use of the principal-value), and $c_1$, $c_2$ and $w$ are integration constants.

Once we have the lapse we can calculate the equations of motion for the metric
\begin{equation}\label{gdotEq}
\dot g_{ij} = \frac{2 N}{\sqrt g} \left( p_{ij}- {\sfrac 1 2} g_{ij} p \right) + \nabla_i \xi_j + \nabla_j \xi_i \,,
\end{equation}
using the spherical symmetry ansatz we get that the $\dot g_{\theta\theta}$ and  $\dot g_{\phi\phi}$  equations completely fix the shift vector:
\begin{equation}\label{SolEquation_gdotrr}
\xi_i = \delta^r{}_i  \left( f \, N + \dot \sigma \right) /\sigma' \,.
\end{equation}
Replacing the above solution of $\xi_i$ in the  $\dot g_{rr}$ equation (as well as the solutions of the ADM constraints), we find that the equation reduces to (with $\dot{\langle p \rangle}$ I mean $\partial_t \langle p \rangle$, \emph{i.e.} the time derivative of the spatial average, and not the other way around)
\begin{equation}\label{pdotEqs}
\begin{aligned}
\left(\langle p \rangle (4\dot A  + 2 c_2) +A  (4 \dot{\langle p \rangle} + w)   - 48 \,\dot m \right) \sigma^{3/2} + \frac{\langle p \rangle}{3} \left(4 \dot{\langle p \rangle}+ w\right) \sigma^{3} + 6 A  \left(2 \dot A + c_2\right)= 0\,.
\end{aligned}
\end{equation}
In order for the above equation to hold for any choice of $\sigma(r)$ the only possibility is that 
\begin{equation}\label{SolutionEqOfMotionThinShell}
\begin{aligned}
&c_2 = -2\,\dot A \,,
&&
w = - 4 \, \dot{\langle p \rangle} \,,
&&
\dot m  = 0 \,.
\end{aligned}
\end{equation}
We got that the Misner--Sharp mass is conserved, and that two of the three integration constants of the lapse are fixed. The third integration constant is arbitrary because it only amounts to a global rescaling of the unit of time.

The equations of motion for the momenta are
\begin{equation}
\begin{aligned}\label{pdotEq}
\dot p^{ij} =& \frac{N}{2\sqrt g} g^{ij} \left( p^{k\ell} p_{k\ell} - {\sfrac 1 2} p^2\right)- \frac{2 N}{\sqrt g} \left( p^{ik} p_k{}^j - {\sfrac 1 2} p \, p^{ij} \right)   - N \sqrt g \left( R^{ij} - {\sfrac 1 2} g^{ij} R + \Lambda g^{ij} \right) 
\\
&+ \nabla_k (p^{ij} \xi^k) - p^{ik} \nabla_k \xi^j - p^{kj} \nabla_k \xi^i + \sqrt g \left( \nabla^i \nabla^j N - g^{ij} \Delta N \right)  \,,
\end{aligned}
\end{equation}
these equations are identically satisfied if one imposes the conditions (\ref{SolutionEqOfMotionThinShell}), and therefore add no further information. We have been able to solve exactly the spherically symmetric ADM-CMC equations in vacuum (with a cosmological constant).

\section{Introduction of matter: thin shells of dust}\label{ThinShellSection}

\subsection{Jump Conditions and Symplectic Potential}

We are now ready to introduce matter. Following~\cite{ThinshellPaper}, we use the simplest form of spherically-symmetric matter: a thin shell of dust.
A shell of dust has only two Hamiltonian degrees of freedom: its coordinate radius $R$ and its radial momentum $P$. Moreover it is characterized by one constant: its rest mass $M$.
The constraints~(\ref{ADMconsts}) are modified in the following way by the addition of a thin shell of dust (see~\cite{ThinshellPaper}):\footnote{The angular integral avoids introducing unnecessary complications such as the pull-back of the metric on a constant-$r$ surface.}
\begin{equation}\label{ADMconsWIthShell}
\begin{aligned}
\!\!\!\!\!\!\!\!& \int \mathcal H \, d\theta d\phi   +  4 \pi \, \delta(r-R) \sqrt{g^{rr} \, P^2 +M^2 }   \approx 0 \,, \\
\!\!\!\!\!\!\!\!& \int  \mathcal H_i \,  d\theta d \phi  +  4 \pi \, \delta^r{}_i \delta (r-R) \, P \approx 0 \,,
\end{aligned}
\end{equation}
while the $\mathcal C \approx 0$ constraint is unchanged.

This modification is localized at the location of the shell, $r=R$, and the equations (\ref{SphericalADMconsts}), (\ref{SphSymmLFE1}), (\ref{pdotEq}), (\ref{gdotEq}), as well as their solutions (\ref{SolConstraints}), (\ref{SolutionLFEtwinshell}), (\ref{SolutionEqOfMotionThinShell})  continue to hold away from the sphere $r=R$.
However, the delta-function introduces a discontinuity in the derivative of the function with the highest derivative in each equation~\cite{ThinshellPaper}, and therefore the integration constants $A$, $m$, $k$, $c_1$, $c_2$ and $w$ will take different values on each side of the shell. Denoting by a subscript `$\cdot_-$' the constants in the region $r<R$, and `$\cdot_+$' those in the region $r>R$, we get that Eqs.~(\ref{ADMconsWIthShell}) imply the following jump conditions (see~\cite{ThinshellPaper}):
\begin{equation}\label{AllJumpconditionsForOneShell}
 A_+ - A_- =  - \frac{\sigma^{\frac 1 2}(R ) }{2\mu(R )} P   \,,
 \qquad
 \lim_{r\to R ^+}  \sigma'(r)  - \lim_{r\to R ^-}  \sigma'(r)   = - {\sfrac 1 2} \sqrt{P ^2 + M ^2 \mu^2(R )} \,.
\end{equation}
Each equation of motion, when extended with the source terms from Eq.~(\ref{ADMconsWIthShell}), leads to a jump condition  of its own. These jump conditions however are not independent, and are automatically satisfied once Eqs.~(\ref{AllJumpconditionsForOneShell}) are~\cite{ThinshellPaper}.

Altogether, the jump conditions make the metric, the lapse and the shift continuous but with discontinuous radial derivative. The momentum is instead discontinuous (but bounded).
These discontinuities depend on the integration constants $A$, $m$ and $k$ taking different values on the two sides of each shell, and they coincide with Israel's junction conditions~\cite{IsraelJunctionConditions}.

In order to discuss the dynamics of the system, we need to know which of the reduced-phase-space variables are canonically conjugate to each other. In other words, we need to calculate the symplectic form. By definition, the conjugate variables of the extended phase space are $g_{ij}$ and $p^{ij}$, as well as $R$ and $P$. Therefore the pre-symplectic potential is
\begin{equation}
\theta = \int_{S^3} p^{ij} \, \delta g_{ij} \,  \d r \d \theta \d \phi  + 4 \pi  \, P  \delta R  \,,
\end{equation}
restricting it through spherical symmetry and integrating in $\d\theta \d\phi$ we get
\begin{equation}
\theta = 4 \pi  \int_0^\pi \d r  \left( 2 f \, \delta \mu + s \, \delta \sigma\right)+ 4 \pi \, P  \delta R  \,.
\end{equation}
Now,  imposing the CMC constraint~$\mu \, f = \mu \, \langle p \rangle \, \sigma - s \, \sigma$, and the solution to the diffeo constraint~(\ref{SolConstraints}),
\begin{equation}
\begin{aligned}
\theta =~& 4 \pi  \int_0^\pi \d r  \left( 2 f \, \delta \mu -\frac{ \mu \, f}{\sigma} \delta \sigma + \langle p \rangle \, \mu \, \delta \sigma \right) + 4 \pi \, P  \delta R   \\
&= - 4 \pi  \int_0^\pi \d r \frac{2 \mu}{\sqrt{\sigma}} \delta (f \sqrt \sigma)+ \langle p \rangle 4 \pi  \int_0^\pi \d r \mu \delta \sigma  + 4 \pi \, P  \delta R   \\
&= - \frac{8 \pi}{3}  \int_0^\pi \d r \,\sigma \, \mu \, \delta \langle p \rangle  + 4 \pi  \, P  \delta R  \\
&-  8 \pi \int_0^\pi  \frac{\mu}{\sqrt \sigma}  \delta \left[ A_-  \, \Theta(R -r) +  A_+ \, \Theta(r-R )  \right]\,,
\end{aligned}
\end{equation}
using the first of the two jump conditions in Eq.~(\ref{AllJumpconditionsForOneShell})
the symplectic potential reduces to
\begin{equation}\label{IsotropicSymplectic0}
\begin{aligned}
\theta &=  - \frac{2}{3} V \delta \langle p \rangle -  8 \pi  \left[ \delta  A_-  \int_0^{R } \d r  \frac{\mu}{\sqrt \sigma}  + \delta  A_+  \int_{R }^{\pi} \d r  \frac{\mu}{\sqrt \sigma} \right]  \,,
\end{aligned}
\end{equation}
where $V = 4 \pi  {\displaystyle \int_0^\pi \sigma \, \mu \, \d r}$ is the on-shell volume.
Now, using the isotropic gauge condition $\mu = \sqrt \sigma/\sin r$,~(\ref{IsotropicSymplectic0}) becomes
\begin{equation}
\begin{aligned}
\theta &=  - \frac{2}{3} V \delta \langle p \rangle  +  8 \pi   ( \delta A_+ - \delta A_-) \log \left(\tan \frac {R} 2 \right)     \,,
\end{aligned}
\end{equation}
and, recalling Eq.~(\ref{AllJumpconditionsForOneShell}), $ A_+ - A_- =  -  {\frac 1 2}  P \sin R   $, we get
\begin{equation}
\theta =  - \frac{2}{3} V \delta \langle p \rangle  - 4 \pi   \log \left(\tan \frac {R} 2 \right)   \delta \left(  P  \sin R  \right)  \,,\end{equation}
which, modulo an exact form, is identical to
\begin{equation}\label{IsotropicSymplectic}
\theta =  - \frac{2}{3} V \delta \langle p \rangle  - 4 \pi R \delta  P  \,.
\end{equation}
Everything we said so far applies identically to more than one shell. The symplectic potential, for example, with many shells turns into $\theta =  - \frac{2}{3} V \delta \langle p \rangle  - 4 \pi  \sum_a R_a \delta P_a$.

\subsection{Single Shell Universe}\label{SubsecSingleShellUniverse}

\begin{figure}[b!]
\floatbox[{\capbeside\thisfloatsetup{capbesideposition={right,center},capbesidewidth=0.37\textwidth}}]{figure}[\FBwidth]
{\caption{\footnotesize The `single shell' universe: the spatial manifold has the topology of the sphere $S^3$ and contains one thin shell  which divides the manifold into  the N and S polar regions. Both regions have to have $A_+=A_-=m_+=m_-=0$ in order for the geometry to be regular at the poles. The shell is characterized by a coordinate-radius degree of freedom $R$ and a radial-momentum degree of freedom $P$,   which are related to the jump in the integration constants $A$, $m$ and $k$. Since $A_+ = A_- = 0$, the momentum of the shell, $P$ is forced to vanish.
}\label{SingleShell_SphereDiagram}}
{\includegraphics[width=0.6\textwidth]{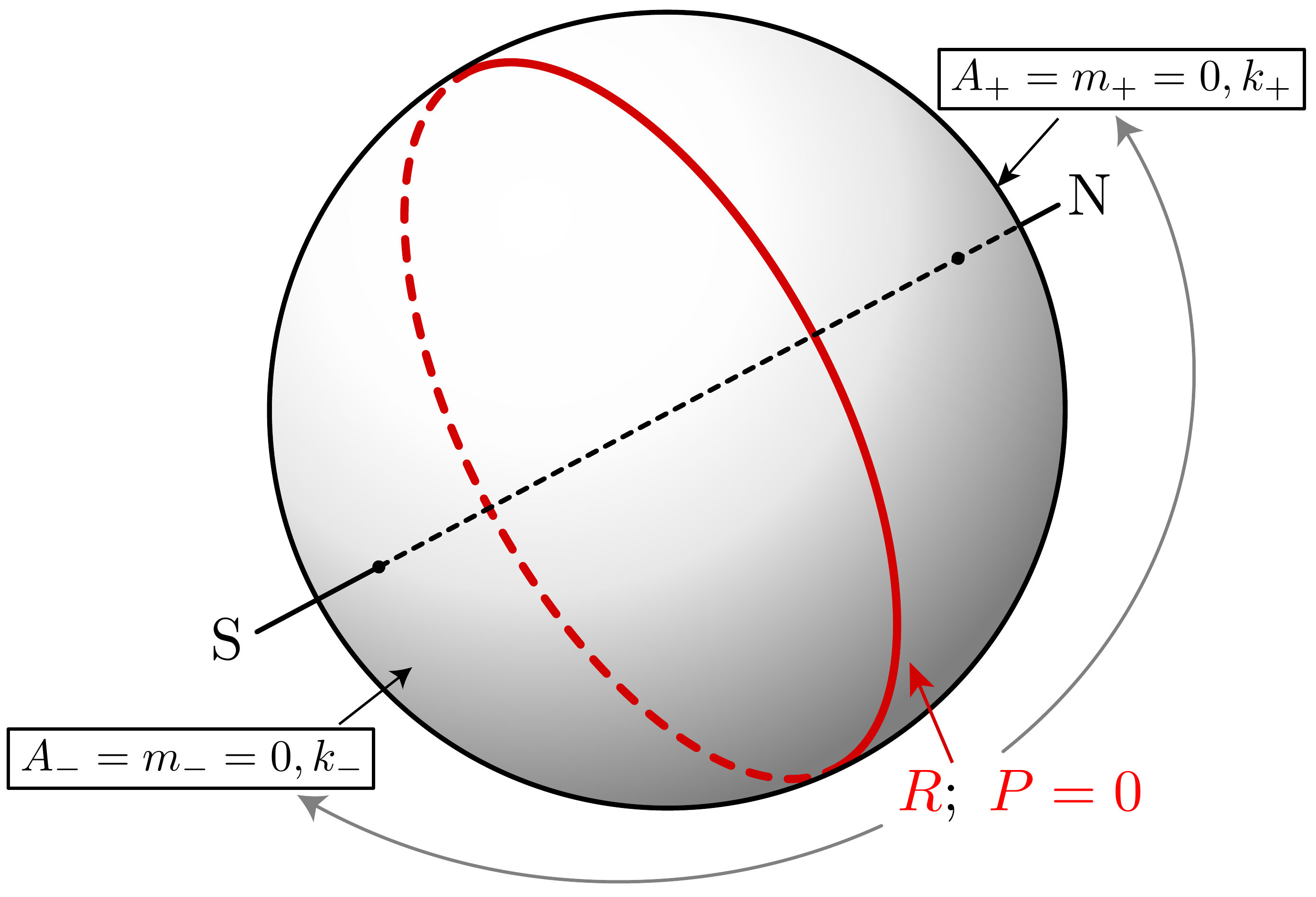}}
\end{figure}

If we include only one shell of dust, the manifold is divided into two regions (we will call them `$+$' and `$-$') which include a pole. Therefore the integration constants $A$ and $m$ are zero in both regions, $A_-=A_+ = m_- = m_+ =0$. Using this in the solution of the constraints, Eq.~(\ref{SolConstraints}),
\begin{equation}
\mu^2(r) = \frac{(\sigma')}{4 \sigma - {\frac 1 9} (12 \Lambda - \langle p\rangle^2) \sigma^2} ~~~ \forall r < R\,, r > R \,.
\end{equation}
Then the continuity of $\mu$ across the shell imposes that $\displaystyle |\lim_{r \to R^+}\sigma'(r)| = |\lim_{r \to R^-} \sigma'(r)|$. Now,  the second of Eqs.~(\ref{AllJumpconditionsForOneShell}) imposes that
\begin{equation}
\lim_{r\to R ^+}  \sigma'(r)  - \lim_{r\to R ^-}  \sigma'(r)   = - {\sfrac 1 2} \sqrt{P ^2 + M ^2 \mu^2(R )} \,,
\end{equation}
while the first sets $P \approx 0$, because $A_+=A_-=0$. So, unless $M=0$, the left and right limits of $\sigma'$ must be equal in magnitude but opposite in sign:
\begin{equation}\label{SigmaDerivativeAtTheShell}
\lim_{r\to R ^+}  \sigma'(r)  = - \lim_{r\to R ^-}  \sigma'(r) =  -  {\sfrac 1 4} \sqrt{P^2 + M ^2  \mu^2(R ) } \approx  - {\sfrac 1 4}  M   \mu (R )  \,.
\end{equation}
Assuming $A_-=A_+ = m_- = m_+ =0$ we can calculate explicitly the metric in isotropic coordinates, such that $\d s^2 =  \mu^2(r) \left[ \d r^2 + \sin^2 r \left( \d \theta^2 + \sin^2 \theta \d \phi^2 \right) \right] $, which implies $\sigma = \sin^2 r \, \mu^2$. This last condition can be considered as a differential equation for $\sigma$:
\begin{equation}\label{SingleShell_IsotropicEquation}
\frac{(\sigma')^2}{4 \sigma - {\sfrac 1 9}  \left( 12 \, \Lambda - \langle p \rangle^2 \right) \sigma^2 } = \frac{\sigma}{\sin^2 r} \,,
\end{equation}
which is solved by
\begin{equation}\label{SingleShellSigma}
\sigma  = \frac{36}{12 \Lambda -\langle p \rangle^2} \left[ 1 - \left( \frac{1- k^2 \tan^2 \frac r 2 }{1 + k^2 \tan^2 \frac r 2} \right)^2 \right] \,.
\end{equation}
In order for the solution above to be positive, we should assume $k$ real if $12 \Lambda -\langle p \rangle^2$, and imaginary otherwise. In other words, the quantity $k^2/(12 \Lambda -\langle p \rangle^2)$ is always positive.

\begin{figure}[t!]
\begin{center}
\includegraphics[height=0.4\textwidth]{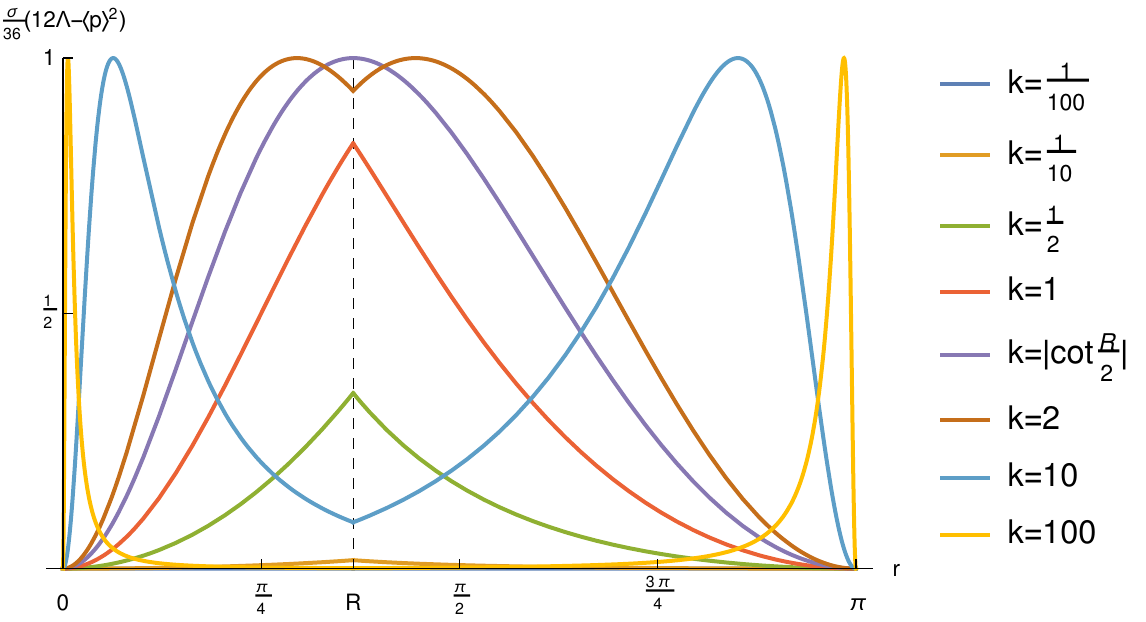}
\caption{\footnotesize
Plot of the solution of Eq.~(\ref{SingleShellSigma2}) for a particular choice of $R $ for various values of $k$.}\label{SingleShellSigmaPlot2}
\end{center}
\end{figure}

As stated in Eq.~(\ref{SigmaDerivativeAtTheShell}) the function $\sigma(r)$ needs to flip sign of its derivative at  $r = R $.  If we call $k_-$$(k_+)$ the value of the integration constant $k$ at the left (right) of the shell, we can impose
\begin{equation}
\lim_{r\to R ^-} \sigma'(k_-,r) 
=
 - \lim_{r\to R ^+}\sigma'(k_+,r)  \,,
\end{equation}
such an equation admits the following solution:
\begin{equation}
k_+ = \frac{\cot^2 \frac {R } 2}{k_-}  \,,
\end{equation}
which implies that the full expression of $\sigma(r)$ at each side of the shell is
\begin{equation}\label{SingleShellSigma2}
\sigma =  {\sfrac{36}{12 \Lambda -\langle p \rangle^2}} \times   \left\{
\begin{array}{ll}
  1 - \left( \frac{1- k^2 \tan^2 \frac r 2 }{1 + k^2 \tan^2 \frac r 2} \right)^2   & \text{\it for } r<R  
\\
  1 - \left( \frac{k^2- \cot^4 \frac {R } 2 \,  \tan^2 \frac r 2 }{k^2  + \cot^4 \frac {R } 2 \, \tan^2 \frac r 2} \right)^2  
& \text{\it for } r>R 
\end{array}\right. \,,
\end{equation}
in Fig.~\ref{SingleShellSigmaPlot2} I plot the function $\sigma$ of Eq.~(\ref{SingleShellSigma2}), divided by $\frac{12 \Lambda -\langle p \rangle^2}{36}$ for a range of values of $k$ and for a particular choice of $R$.  
We then have a 1-parameter family of metrics which are exact solutions of the local parts of the constraints. All that is left over to solve are the jump conditions. The diffeomorphism one simply reduces to the constraint $P \approx 0$, while the Hamiltonian one takes a more complicated functional form:
\begin{equation}
\begin{aligned}
h(k,R ,\langle p \rangle )  &= 
\lim_{r\to R ^+} \sigma'(r)  - \lim_{r\to R ^-} \sigma'(r)  + {\sfrac 1 2}    \sqrt{P ^2 + M ^2  \sigma(R ) \sin^{-2} R  }
\\
&=
\frac{8 \, k^2 \cos \left(\frac{R}{2}\right) \left[ k^2-\cot ^2\left(\frac{R}{2}\right) \right]}{\sin ^3\left(\frac{R}{2}\right) \left[ k^2+\cot ^2\left(\frac{R}{2}\right)\right]^3}  + {\sfrac 1 2}    \sqrt{P ^2 + M ^2  \sigma(R ) \sin^{-2} R  } \approx 0 \,.
\end{aligned}
\end{equation}
We are left with a 4-dimensional phase space, $(R,P,k,\langle p \rangle)$ and two constraints: $P\approx0$, $h \approx 0$. We need to perform a Dirac analysis and check whether the constraints are first- or second-class. To do so we need to calculate the symplectic form. Recall from the previous Section that the symplectic potential in isotropic gauge is $\theta =  - \frac{2}{3} V \delta \langle p \rangle  - 4 \pi R  \delta  P $. The volume $V$ is a function of $R $, $k$ and $\langle p \rangle$:
\begin{equation}\label{SingleShellExactVolume}
\begin{aligned}
&V= 4 \pi \int \d r \sigma \mu = 8 \pi \int_0^{\sigma(R )}  \frac{\sigma \d\sigma}{\sqrt{4 \sigma - {\sfrac 1 9}  \left( 12 \, \Lambda - \langle p \rangle^2 \right) \sigma^2}}\\
&=\textstyle
\frac{1728 \, \pi}{\left(12 \Lambda -\langle p \rangle^2\right)^{3/2}}  \left\{ \tan^{-1} \left[ k \tan \left(\frac{R }{2}\right)\right] - \frac{k \tan \left(\frac{R}{2}\right) \left(1-k^2 \tan ^2\left(\frac{R}{2}\right)\right)}{\left(k^2 \tan ^2\left(\frac{R}{2}\right)+1\right)^2}\right\} \,.
\end{aligned}
\end{equation}
The variation of the volume takes a simple form:
\begin{equation}
\delta V = \frac{8 k^2 \sin ^2 R (\sin R \, \delta k +k \, \delta R)}{\left(1 +k^2+(1 -k^2)  \cos R\right)^3} \,,
\end{equation}
so the symplectic potential is simply
\begin{equation}
\omega = \delta \theta =   - \frac{2}{3} w(R,k) \left( \sin R \, \delta k \wedge \delta \langle p \rangle +k \, \delta R \wedge \delta \langle p \rangle \right)  - 4 \pi \delta R \wedge  \delta  P \,,
\end{equation}
where $w(R,k) = \frac{8 k^2 \sin ^2 R}{\left(1 +k^2+(1 -k^2)  \cos R\right)^3}$.
Now, the Poisson brackets between any two phase-space functions are given by the inverse of the symplectic form:
\begin{equation}
\{ f , g \} = \partial_i f \, (\omega^{-1})^{ij} \, \partial_j g \,, 
\end{equation}
and since

\begin{equation}\label{SingleShellSymplectic2-Form-Matrix}
\begin{aligned}
\omega_{ij} = \frac{1}{2} \left(
\begin{array}{cccc}
 0 & -\frac{2}{3} w \,  \sin R  & 0 & 0 \\
 \frac{2}{3} w \,  \sin R & 0 & \frac{2}{3} w \, k & 0 \\
 0 & - \frac{2}{3} w \, k & 0 & -4 \pi  \\
 0 & 0 & 4 \pi  & 0 \\
\end{array}
\right) \,,
\end{aligned}
\end{equation}
(where $i,j= \left(  k ,  \langle p \rangle ,  R  ,  P  \right)$, the inverse can be shown to be
\begin{equation}\label{SingleShellSymplectic2-Form-InverseMatrix}
\begin{aligned}
 (\omega^{-1})^{ij}  = \left(
\begin{array}{cccc}
 0  & \frac 3 {w \,  \sin R } & 0 & -  \frac 1 {2\pi} \frac{k}{\sin R} \\
 - \frac 3 {w \,  \sin R} & 0 & 0 & 0 \\
 0  & 0 & 0 &    \frac 1 {2\pi} \\
   \frac 1 {2\pi} \frac{k}{\sin R}   & 0 & -   \frac 1 {2\pi} & 0 
 \end{array}
\right) \,.
\end{aligned}
\end{equation}
Then the Poisson brackets take the following explicit form:
\begin{equation}
\{ f , g \}  = \textstyle \frac 1 {w \,  \sin R} \left[ 3 \left(\frac{\partial f}{\partial k}\frac{\partial g}{\partial \langle p\rangle }-\frac{\partial f}{\partial \langle p\rangle }\frac{\partial g}{\partial k}\right) 
-\frac{1}{2 \pi } w \, k
\left(\frac{\partial f}{\partial k}\frac{\partial g}{\partial P }-\frac{\partial f}{\partial P }\frac{\partial g}{\partial k}\right) \\\right]+\frac{1}{2 \pi } \left(\frac{\partial f}{\partial R }\frac{\partial g}{\partial P }-\frac{\partial f}{\partial P }\frac{\partial g}{\partial R }\right) \,.
\end{equation}
The Poisson brackets between $h(k,R ,\langle p \rangle ) $ and $P $ then is:
\begin{equation}
\{ h , P  \} =
-\frac{1}{2 \pi } \frac{k}{\sin R} \, \frac{\partial h}{\partial k}   +\frac{1}{2 \pi } \, \frac{\partial h}{\partial R } \,,
\end{equation}
and an explicit calculation reveals that
\begin{equation}
\{ h , P  \} \approx - \frac{\cot R }{2 \pi} h \approx 0 \,,
\end{equation}
so the two constraints are first-class.

We have a four-dimensional phase space with two first-class constraints. One linear combination of the constraints can be interpreted as generating gauge transformations and indicating an unphysical degree of freedom, but the other linearly independent one cannot (see~\cite{FlaviosSDtutorial}, in particular `Barbour and Foster's exception to Dirac's theorem), because it plays the role of Hamiltonian constraint generating the dynamics. It is convenient to take $P  \approx 0$ as the gauge constraint, which can be gauge-fixed with 
\begin{equation}
\chi = R  - \bar R \approx 0 \,,
\end{equation}
where $\bar R \in (0 , \pi)$ is any function of time (the simplest choice is a constant). $\chi$ is trivially first-class with respect to $h$ and second-class with respect to $P $. 
Replacing the gauge constraint $P  \approx 0$ and the gauge fixing $\chi  \approx 0 $ in the leftover Hamiltonian constraint $h\approx 0$ we get
\begin{equation}
h \propto \frac{M}{\sqrt{2 \cos^4 {\sfrac{\bar R}2}  \left(k^2 \tan ^2{\sfrac{\bar R}2}+1\right)^2}}
-
 \sqrt{\frac{36 \, k^2}{12 \Lambda - \langle p \rangle^2}}  \frac{  8 \, \cos {\sfrac{\bar R}2} \left( k^2-\cot ^2{\sfrac{\bar R}2}\right)}{\sin^3 {\sfrac{\bar R}2} \left( k^2+\cot ^2{\sfrac{\bar R}2}\right)^3}
\approx 0 \,,
\end{equation}

Assuming that  $12 \Lambda > \langle p \rangle^2$, we can take $k$ real and positive, and the above equation  is equivalent to the following sixth-order polynomial in $k$:
\begin{equation}\label{k_polynomial}
\left(k^2+\cot ^2{\sfrac{\bar R}2}\right) \left(k^4 m-96 k^3 \cot {\sfrac{\bar R}2}+2 k^2 m \cot ^2{\sfrac{\bar R}2}+96 k \cot ^3{\sfrac{\bar R}2}+m \cot ^4{\sfrac{\bar R}2}\right)= 0 \,,
\end{equation}
where $m = M \sqrt{12 \Lambda - \langle p \rangle^2}$.

\begin{figure}[ht!]
\begin{center}
\includegraphics[height=0.28\textheight]{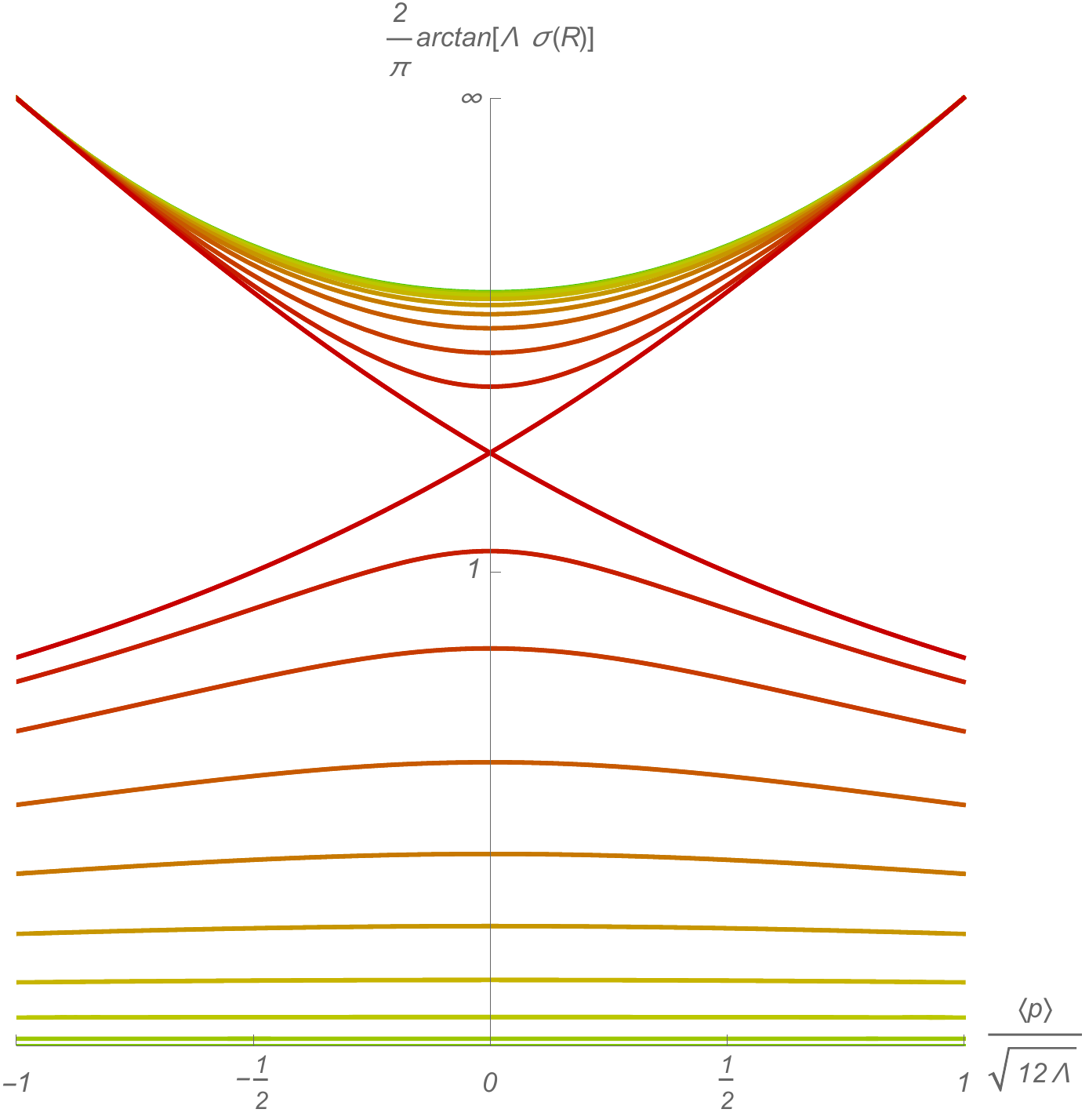}~~\includegraphics[height=0.28\textheight]{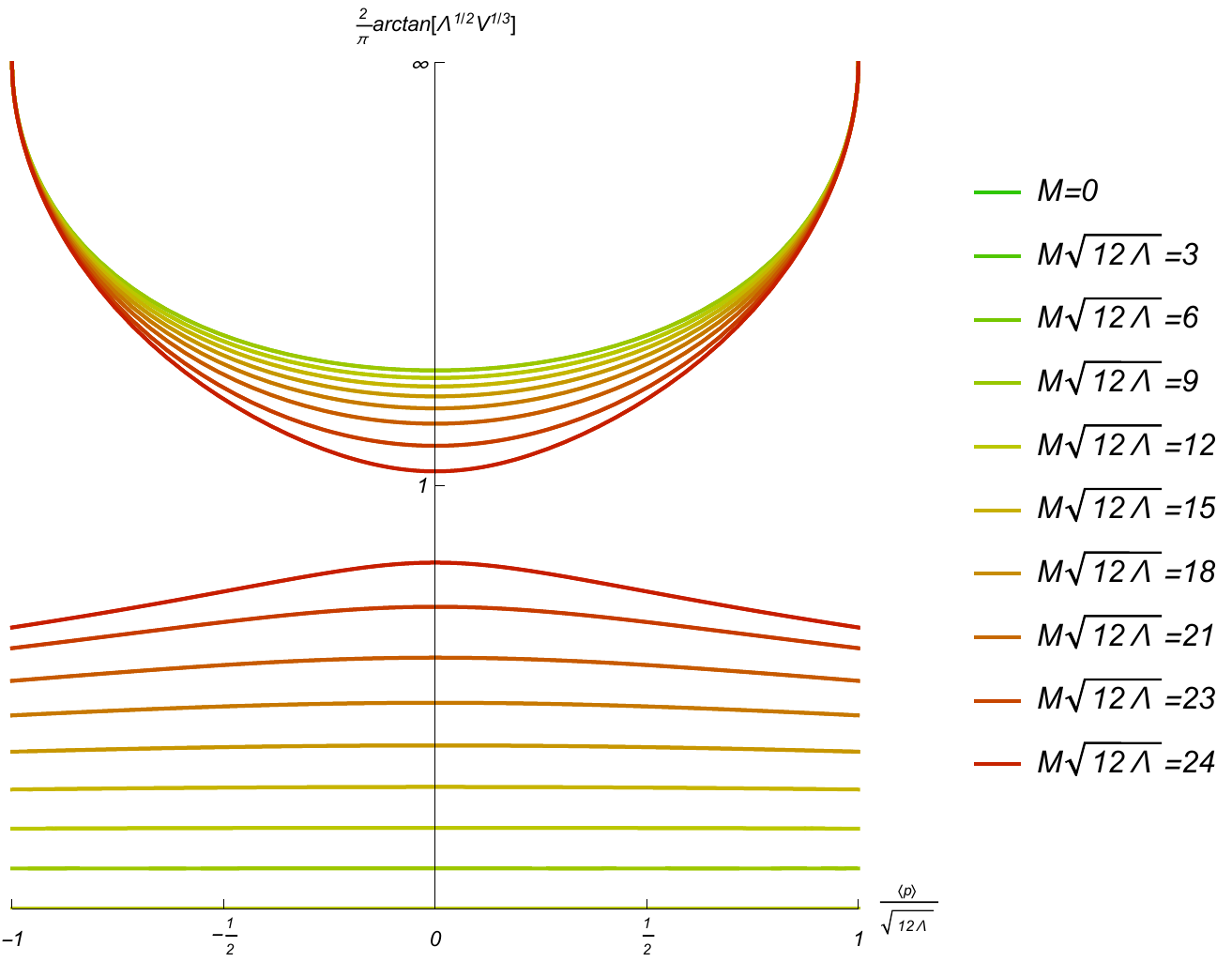}
\caption{\footnotesize
Plot of $\sigma(R)$ (left) and $V$ (right) as functions of York time $\langle p \rangle$ for a set of values of the rest mass. The vertical axis has been compactified by taking the $\arctan$.
As we can see, for each choice of the dimentionless parameter $M \sqrt{12 \Lambda}$ there are two conjugate solutions: one which goes from infinite volume/areal radius at the shell (at $\langle p \rangle = - \sqrt{12\Lambda}$) to a finite minimum (at $\langle p \rangle = 0$) and back to infinity (at $\langle p \rangle = +\sqrt{12\Lambda}$). Another one that goes from a finite volume/areal radius (at $\langle p \rangle = \pm \sqrt{12\Lambda}$) to a finite maximim (at $\langle p \rangle = 0$). In the zero-rest-mass limit the first kind of solutions tend to an acceptable result: two compact patches of de Sitter universe glued at their border, which is delimited by a lightlike shell. The second kind stops making sense as $M \to 0$: both the areal radius $\sigma(R)$ and the volume are zero throughout the  solution. In the opposite limit, $M \to \frac{24}{\sqrt{12 \Lambda}}$, the solution ceases to be smooth, because the first and second kind of solutions meet at a point at $\langle p \rangle =0$, and $d \sigma(R)/d\langle p\rangle$ has a discontinuity at that point. This is a signal that at that point the `cosmological horizon scale' associated to the cosmological constant and the `Schwarzschild horizon scale' associated to the mass-energy of the shell coincide. The physics of this family of solutions that I uncovered will be investigated in future works.
}\label{SingleShellplots}
\end{center}
\end{figure}

We can assume that $k^2 \neq - \cot ^2{\sfrac{\bar R}2}$ because $k \in \mathbbm{R}$ (otherwise $\sigma$ would be negative), so Eq.~(\ref{k_polynomial}) is equivalent to a fourth-order equation. Its discriminant is proportional to 
$$ m^2 \left(m^2-576\right)^2 \sin ^{20}(\bar R) \csc^8\left(\frac{\bar R}{2}\right) \,, 
$$
which is always positive. Therefore there are either four or zero real roots.
The former case holds only if both $\left(m^2-1728\right) \sin ^4\left(\frac{\bar R}{2}\right) \sin ^2(\bar R)  <0$ and $\left(m^2-576\right) \sin ^8\left(\frac{\bar R}{2}\right) \sin ^4(\bar R)<0$. So, in order for real roots to exist, we have to have $m^2 < 576$, that is,
\begin{equation}
M^2  < \frac{24^2}{12 \Lambda - \langle p \rangle^2} \,.
\end{equation}
In summary, we found that the dynamics of the single-shell universe is completely trivial: the radial coordinate of the shell, $R$ is unphysical (even in isotropic gauge), because its conjugate momentum $P$ is a first-class constraint. All we can do is to impose $P\approx 0$ in $h(k,R,\langle p \rangle) \approx 0$ and we get a functional relation between $k$ and $\langle p \rangle$. For a given value of the rest mass $M$ this completely fixes the CMC metric as a function of the York time $\langle p \rangle$. In Fig.~\ref{SingleShellplots} I plot $\sigma(R)$ and the volume $V$ from Eq.~(\ref{SingleShellExactVolume}) as functions of York time $\langle p \rangle$, for a set of choices of $M$ between $0$ and the maximum $24/\sqrt{12 \Lambda - \langle p \rangle^2}$.

We were able to solve analytically  the single-shell universe, because in this case the isotropic gauge condition leads to an equation for $\sigma$, (\ref{SingleShell_IsotropicEquation}), that can be solved exactly. However this will be a luxury that we cannot afford in the following section, and we need to be prepared to study the dynamics even when an explicit solution of the Hamiltonian constraint in isotropic gauge is not available. For this reason, in preparation to the next section, I here calculate again the symplectic potential without assuming any particular radial gauge. I will instead try to exploit as much as I can all the gauge-independent information we have about the form of the solution of the constraints.
To do so, I need to make only reference to the variables $\rho = \sqrt{\sigma (R)}$ (the areal radius of the shell), $A_+$ and $A_-$, which do not depend on the radial gauge (as opposed to $R$ and $P$, which, being related to a coordinate system, take a meaning only when a radial gauge is fixed).

Let's begin with the form~(\ref{IsotropicSymplectic0}) for the pre-symplectic potential:
\begin{equation}
\theta =  - \frac{2}{3} V \delta \langle p \rangle -  8 \pi  \left[ \delta  A_-  \int_0^{R } \d r  \frac{\mu}{\sqrt \sigma}  + \delta  A_+  \int_{R }^{\pi} \d r  \frac{\mu}{\sqrt \sigma} \right]  \,,
\end{equation}
introducing the theta functions
\begin{equation}
\Theta_+ (r) = \Theta(r-R) \,, \qquad  \Theta_- (r) = \Theta(R-r) \,, 
\end{equation}
we can write the potential as
\begin{equation}
\begin{aligned}
\theta 
=&  -  8 \pi    \sum_{\beta \in \{ +,-\}}  \int_{\pi}^{0} \frac{\left(\frac 1 3 \sigma^{3/2}(r) \, \delta \langle p \rangle  + \delta A_\beta  \right) \Theta_\beta(r) \, |\sigma'| \d r}{\sqrt{ A^2_\beta  + \left( {\sfrac 2 3} \langle p \rangle  A_\beta - 8 \, m_\beta  \right) \sigma^{3/2}  + 4 \, \sigma^2 - {\sfrac 1 9}  \left( 12 \, \Lambda - \langle p \rangle^2 \right) \sigma^3 }}
\\
=&  -  8 \pi    \int_0^{\sigma(R)} \left( \frac{\partial F_- [A_-,\langle p \rangle ,\sigma]}{\partial \langle p \rangle}  \delta \langle p \rangle  +  \frac{\partial F_- [A_-,\langle p \rangle ,\sigma]}{\partial A_-}  \delta A_- \right) d \sigma 
\\
&-  8 \pi    \int_0^{\sigma(R)} \left( \frac{\partial F_+ [A_+,\langle p \rangle ,\sigma]}{\partial \langle p \rangle}  \delta \langle p \rangle  +  \frac{\partial F_+ [A_+,\langle p \rangle ,\sigma]}{\partial A_+}  \delta A_+ \right) \d \sigma 
   \,,
\end{aligned}
\end{equation}
where
\begin{equation}
F_\beta  =  \log \left(A_\beta + {\sfrac 1 3} \langle p \rangle \sigma^{3/2}   \sqrt{ A^2_\beta  + \left( {\sfrac 2 3} \langle p \rangle  A_\beta - 8 \, m_\beta  \right) \sigma^{3/2}  + 4 \, \sigma^2 - {\sfrac 1 9}  \left( 12 \, \Lambda - \langle p \rangle^2 \right) \sigma^3 } \right) \,.
\end{equation}
Then the symplectic form is
\begin{equation}
\begin{aligned}
\omega = \delta \theta =  -  8 \pi    \bigg{(}&
  \frac{\partial F_- [A_-,\langle p \rangle ,\sigma(R)]}{\partial \langle p \rangle}  \delta \langle p \rangle  +  \frac{\partial F_- [A_-,\langle p \rangle ,\sigma(R)]}{\partial A_-}  \delta A_-   \\
&+ \frac{\partial F_+ [A_+,\langle p \rangle ,\sigma(R)]}{\partial \langle p \rangle}  \delta \langle p \rangle  +  \frac{\partial F_+ [A_+,\langle p \rangle ,\sigma(R)]}{\partial A_+}  \delta A_+ \bigg{)} \wedge \delta \sigma(R)
   \,.
\end{aligned}
\end{equation}
applying $A_\pm = m_\pm =0$
\begin{equation}
\omega  =  -  \frac{16 \pi}{3}    \left( \frac{\sigma^{3/2}(R)}{\sqrt{ 4 \, \sigma^2(R) - {\sfrac 1 9}  \left( 12 \, \Lambda - \langle p \rangle^2 \right) \sigma^3(R)} } \right)   \delta \langle p \rangle  \wedge \delta \sigma(R)
   \,.
\end{equation}
The above equation is essentially stating that the variables $\sigma(R)$ and $\langle p \rangle$ are canonically conjugate.

Let's now discuss the constraints imposed by the jump conditions. The diffeomorphism jump condition [the first of Eqs.~(\ref{AllJumpconditionsForOneShell})] is now just a definition of $P$, which is not a dynamical variable anymore. The Hamiltonian jump condition [the second of Eqs.~(\ref{AllJumpconditionsForOneShell})] can be written in a way that depends only on $\rho$, $A_+$ and $A_-$. In fact, define $\kappa = \lim_{r\to R^-} \sigma'(r)$ and $\gamma = \lim_{r\to R^-} \sigma'(r)$. Using the first of Eqs.~(\ref{AllJumpconditionsForOneShell}) into the second one:
\begin{equation}
\gamma - \kappa = - {\sfrac 1 2} \sqrt{4 \frac{(A_+-A_-)^2}{\rho^2} \mu^2(R) + M^2 \mu^2(R)} \,,
\end{equation}
and dividing by $|\mu(R)|$
\begin{equation}
\frac{\gamma}{|\mu(R)|} - \frac{\kappa}{|\mu(R)|} = - {\sfrac 1 2} \sqrt{4 \frac{(A_+-A_-)^2}{\rho^2}  + M^2} \,,
\end{equation}
we can square the above equation and reorder
\begin{equation}
\frac{\gamma^2}{\mu^2(R)} + \frac{\kappa^2}{\mu^2(R)} - \frac{(A_+-A_-)^2}{\rho^2}  - {\sfrac 1 4} M^2  = 2 \frac{\gamma \, \kappa }{\mu^2(R)}    \,,
\end{equation}
and taking another square
\begin{equation}
\left( \frac{\gamma^2}{\mu^2(R)} + \frac{\kappa^2}{\mu^2(R)} - \frac{(A_+-A_-)^2}{\rho^2}  - {\sfrac 1 4} M^2 \right)^2  = 2 \frac{\gamma^2}{\mu^2(R)} \frac{\kappa^2}{\mu^2(R)}     \,,
\end{equation}
the equation only depends on  $\frac{\gamma^2}{\mu^2(R)} =\frac{\left(\lim_{r\to R^-} \sigma'(r)\right)^2}{\mu^2(R)}  $ and $\frac{\kappa^2}{\mu^2(R)} =\frac{\left(\lim_{r\to R^+} \sigma'(r)\right)^2}{\mu^2(R)}  $. Now we can use the last of Eqs.~(\ref{SolConstraints}) to get rid of $\kappa/\mu^2(R)$ and $\gamma/\mu^2(R)$:
\begin{equation}
\begin{aligned}
\frac{\gamma^2}{\mu^2(R)} =\frac{\left(\lim_{r\to R^-} \sigma'(r)\right)^2}{\mu^2(R)} = \frac{A_-^2}{\rho} + \left( {\sfrac 2 3} \langle p \rangle A_- - 8 m_- \right) \rho + 4 \rho^2 - {\sfrac 1 9} \left( 12 \Lambda - \langle p \rangle^2 \right) \rho^4  \,,
\\
\frac{\kappa^2}{\mu^2(R)} =\frac{\left(\lim_{r\to R^+} \sigma'(r)\right)^2}{\mu^2(R)}  = \frac{A_+^2}{\rho} + \left( {\sfrac 2 3} \langle p \rangle A_+ - 8 m_+ \right) \rho + 4 \rho^2 - {\sfrac 1 9} \left( 12 \Lambda - \langle p \rangle^2 \right) \rho^4  \,,
\end{aligned}
\end{equation}
and we have our contraint purely in terms of $\rho$, $A_+$ and $A_-$. Now we can use the boundary conditions at the poles, $A_\pm = m_\pm =0$, and the constraint simplifies to:
\begin{equation}\label{SingleShellOnshellConstraint}
\frac{M^2}{16} +   {\sfrac 1 9} \left( 12 \Lambda - \langle p \rangle^2 \right) \rho^4 - 4 \rho^2  =0 \,.
\end{equation}
The above constraint admits a real positive $\rho$ only when $M^2 (12 \Lambda - \langle p \rangle^2)<24^2$, which is the same upper bound on the mass that we found above.
Moreover, if we plot the solutions of~(\ref{SingleShellOnshellConstraint}) w.r.t. $\rho^2$ as functions of $\langle p \rangle$ we obtain the same diagram as the left one in Fig.~\ref{SingleShellplots}. We were therefore able to extract the same amount of information as before, but without having to fix the radial gauge.

We were able to solve every aspect of the `single-shell universe' analytically. The result is a system whose dynamics is completely trivial: the coordinate position of the shell $R$ is a gauge degree of freedom (the diffeomorphism constraint reduces to $P \approx 0$ which implies that the conjugate variable, $R$, is a gauge direction). The gauge-invariant degrees of freedom are all completely constrained: once we specify the rest mass of the shell $M$ in units of the cosmological constant $\Lambda$ the evolution is completely fixed and admits no integration constants: there are no adjustable parameters that we can choose to set initial data. The space of solutions is just a point.
This system is, therefore, too trivial for our purposes. We need to add degrees of freedom in order to have a nontrivial solution space.

\newpage

\subsection{`Twin Shell' Universe}\label{SubsecTwinShellUniverse}

\begin{figure}[b!]
\floatbox[{\capbeside\thisfloatsetup{capbesideposition={left,center},capbesidewidth=0.37\textwidth}}]{figure}[\FBwidth]
{\caption{\footnotesize The `twin shell' universe: the spatial manifold has the topology of the sphere $S^3$ and contains two concentric thin shells, which divide the manifold into three regions: the N and S polar regions, and the B region in between (belt). The shell closer to the north (south) pole will be indicated as the N (S) shell. Each region will have different values of the integration constants $A$, $m$ and $k$, but regularity demands that $A_\st{N}=A_\st{S}=m_\st{N}=m_\st{S}=0$. Moreover the two shells will come equipped with a coordinate-radius degree of freedom $R_\st{S}$, $R_\st{N}$ and a radial-momentum degree of freedom $P_\st{S}$, $P_\st{N}$, which will be related to the jump in the integration constants $A$, $m$ and $k$.
}\label{TwinShellPicture}}
{\includegraphics[width=0.6\textwidth]{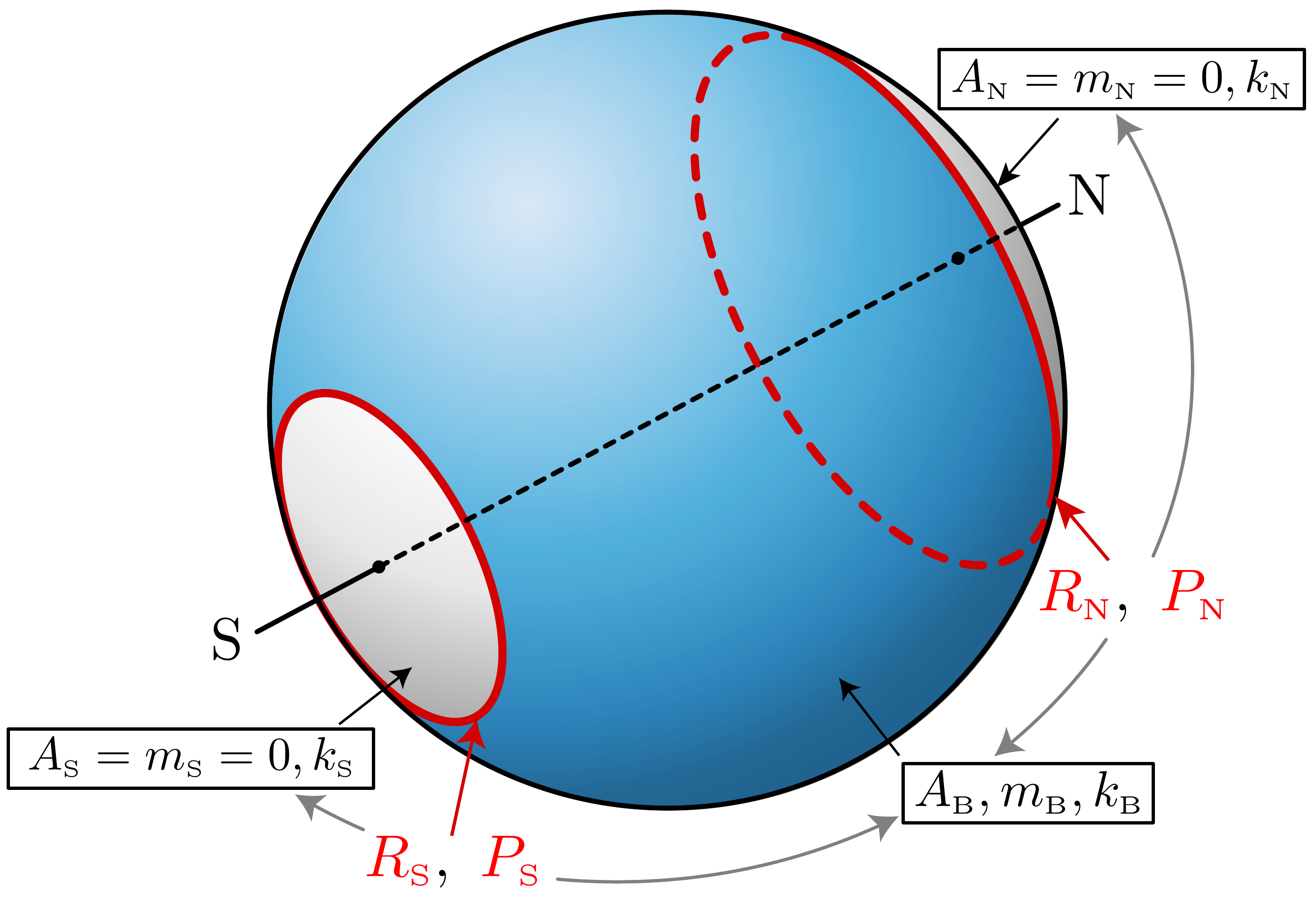}}
\end{figure}

The minimal number of shells we need in order to have a nontrivial dynamics in a compact universe is two.  In fact, if we want the regions around the poles (which we will call `north' and indicate with the subscript `$\cdot_\st{N}$' and `south', indicated with `$\cdot_\st{S}$') to be compact and regular, we need the parameters $A_\st{N}$, $A_\st{S}$, $m_\st{N}$ and  $m_\st{S}$ to be zero (see~\cite{BirkhoffFlavio} and Appendix~~\ref{AppendixBoundaryConditions} for a proof). Then, we can see that one single shell would be dynamically trivial, because in that case from Eq.~(\ref{AllJumpconditionsForOneShell}) $ A_\st{N} - A_\st{S} =  - \sqrt{\sigma(R)} P /[2\mu(R)]=0$, and the single shell would always have zero momentum.  So we introduce two shells, which we will call north and south according to which pole they surround, and indicate with the corresponding subscript. The region in between the shells will be called `belt' and  indicated with `$\cdot_\st{B}$'. See Fig.~\ref{TwinShellPicture} for a diagram of the regions in our manifold.

With two shells, the first of the two jump conditions~(\ref{AllJumpconditionsForOneShell}) translates into
\begin{equation}\label{Twinshell_DeltaAjumpcondition}
A_\st{B} - A_\st{S} = - \frac{\sqrt{\sigma(R_\st{S})}}{2\mu(R_\st{S})} P_\st{S} \,,
\qquad
A_\st{N} - A_\st{B} = - \frac{\sqrt{\sigma(R_\st{N})}}{2\mu(R_\st{N})} P_\st{N} \,.
\end{equation}
Now, calling  $\sigma(R_\st{S}) = \rho^2_\st{S}$, $\sigma(R_\st{N}) = \rho^2_\st{S}$, and the left- and right- derivatives of $\sigma$ at the shells:
\begin{equation}
\gamma_\st{S,N} = \lim_{r\to R_\st{S,N}^+} \sigma'(r) \,,
~~~~
\displaystyle \kappa_\st{S,N} = \lim_{r\to R_\st{S,N}^-} \sigma'(r) \,,
\end{equation}
the second of the jump conditions~(\ref{AllJumpconditionsForOneShell}) reads
\begin{equation}\label{Twinshell_JumpGammaKappa}
\gamma_\st{S}  - \kappa_\st{S}   = - {\sfrac 1 2} \sqrt{P_\st{S}^2 + M_\st{S}^2  \mu^2(R_\st{S}) } \,,
~~~
\gamma_\st{N}  - \kappa_\st{N}   = - {\sfrac 1 2} \sqrt{P_\st{N}^2 + M_\st{N}^2  \mu^2(R_\st{N}) } \,.
\end{equation}
Using Eq.~(\ref{Twinshell_DeltaAjumpcondition}) into Eqs.~(\ref{Twinshell_JumpGammaKappa}): 
\begin{equation}\label{Twinshell_JumpGammaKappa2}\textstyle
\frac{\kappa_\st{S}}{| \mu(R_\st{S}) |}  - \frac{\gamma_\st{S}}{| \mu(R_\st{S}) |}  =  \sqrt{\frac{(A_\st{S} - A_\st{B})^2}{\rho_\st{S}^2} + {\sfrac 1 4} M_\st{S}^2 } \,, \qquad
\textstyle
\frac{\kappa_\st{N}}{| \mu(R_\st{S}) |}   - \frac{\gamma_\st{N}}{| \mu(R_\st{N}) |}    = \sqrt{\frac{(A_\st{B} - A_\st{N})^2}{\rho_\st{N}^2} + {\sfrac 1 4} M_\st{N}^2 } \,.
\end{equation}
by taking twice the square of the above equations, we can make them independent of the signs of $\kappa_{S,N}$ and $\gamma_\st{S,N}$ (the following is a pair of identical equations, in which the subscript $a$ of the quantities $\kappa$, $\gamma$, $A$, $R$ and $M$ can either be $S$ or $N$),
\begin{equation}\label{Twinshell_JumpGammaKappa3}
\begin{aligned}\textstyle
\left(\frac{\kappa_\st{S,N}^2}{\mu^2(R_a)}  + \frac{\gamma_a^2}{\mu^2(R_a)} -   \frac{(A_a - A_\st{B})^2}{\rho_a^2} -  {\sfrac 1 4} M_a^2 \right)^2 = 4 \frac{\kappa_a^2}{\mu^2(R_a)} \frac{\gamma_a^2}{\mu^2(R_a)}\,,
\end{aligned}
\end{equation}
now, using the solution for $\mu(r)$ from Eq.~(\ref{SolConstraints}) at $r= R_\st{S}$ and $r = R_\st{S}$,
{\medmuskip=0mu
\thinmuskip=0mu
\thickmuskip=0mu
\begin{equation}
\begin{aligned}
&\textstyle
\frac{\gamma_\st{S}^2}{4 \mu^2(R_\st{S})}  = \left(\frac {A_\st{S} + \frac 1 3 \langle p \rangle \rho_\st{S}^3 }{2 \rho_\st{S}}\right)^2  - 2 m_\st{S} \, \rho_\st{S} +  \rho_\st{S}^2  - \frac{\Lambda  \rho_\st{S}^4}{3} \,,
&
&\textstyle\frac{\kappa_\st{S}^2}{4 \mu^2(R_\st{S})} = \left(\frac {A_\st{B} + \frac 1 3 \langle p \rangle \rho_\st{S}^3 } {2\rho_\st{S}}\right)^2  - 2 m_\st{B} \, \rho_\st{S} +  \rho_\st{S}^2  - \frac{\Lambda  \rho_\st{S}^4}{3} \,,
\\
&\textstyle
\frac{\gamma_\st{N}^2}{4\mu^2(R_\st{N})} = \left(\frac {A_\st{B} + \frac 1 3 \langle p \rangle \rho_\st{N}^3 } {2\rho_\st{N}}\right)^2  - 2 m_\st{B} \, \rho_\st{N} + \rho_\st{N}^2  - \frac{\Lambda  \rho_\st{N}^4}{3} \,,
&
&\textstyle\frac{\kappa_\st{N}^2}{4\mu^2(R_\st{S})} = \left(\frac {A_\st{N} + \frac 1 3 \langle p \rangle \rho_\st{N}^3 } {2\rho_\st{N}}\right)^2  - 2 m_\st{N} \, \rho_\st{N} +  \rho_\st{N}^{2}  - \frac{ \Lambda  \rho_\st{N}^4}{3} \,,
\end{aligned}\label{EqsForGammasAndKappas}
\end{equation}}
where $\rho_a = \sqrt{\sigma(R_s)}$ are the areal radii at the two shells, and recalling that, in order to keep the poles compact and smooth we need to have $A_\st{S} = A_\st{N}=0$ and $m_\st{S}=m_\st{N}=0$, we end up with the following two on-shell conditions:
\begin{equation}\label{Twinshell_OnshellCondition}
\frac{M_a^4}{16}+4 A_\st{B}^2 \left(T \rho _a^2-4\right)+M_a^2 \rho _a \left(T \rho _a^3 - 4 \rho _a-2 X\right)+16 X^2 \rho _a^2 = 0\,,
\end{equation}
where
\begin{equation}
T = {\sfrac 1 9} \left( 12 \Lambda - \langle p \rangle^2 \right)\,,
\qquad
X = {\sfrac 1 6} \langle p \rangle A_\st{B} - 2 \, m_\st{B} \,.
\end{equation}
Conditions~(\ref{Twinshell_OnshellCondition}) are two identical equations 
involving the same $A_\st{B}$ and two different areal radii $\rho_a=(\rho_\st{S},\rho_\st{N})$ and rest-masses $M_a=(M_\st{S},M_\st{N})$. By rescaling both equations with appropriate powers of $m_\st{B}$ we can make them dimensionless. This requires introducing dimensionless variables analogue to those of Eq.~(\ref{MordorPoly}):
\begin{equation}
C = \frac{A_\st{B}}{2 \, m_\st{B}^2} \,, ~~~   \tau = | m_\st{B} |\, \langle p \rangle \,, ~~~ \lambda = m_\st{B}^2 \,\Lambda \,, ~~~ z_a = \frac{\rho_a}{|m_\st{B}|} \,, ~~~ M_a = |m_\st{B}| \, \mu_a \,.
\end{equation}
Then the two Equations~(\ref{Twinshell_OnshellCondition}) can be written
\begin{equation}\label{Twinshell_OnshellCondition_dimensionless}
\frac{\mu_a^4}{16}+ \mu_a^2 z_a \left[ \pm 4  -   {\sfrac 2 3} C \tau -{\sfrac 1 9} z_a^3 \left(\tau^2-12 \lambda \right)- 4 z_a\right] - \frac{64}{3} \left[\pm C \tau  z_a^2 -  C^2 \left(\lambda  z_a^2-3\right)�- 3 z_a^2 \right] = 0 \,,
\end{equation}
where the sign $+$ corresponds to $m_\st{B} >0$ and $-$ corresponds to $m_\st{B} <0$.

\begin{figure}[b!]
\center
\includegraphics[width=0.3\textwidth]{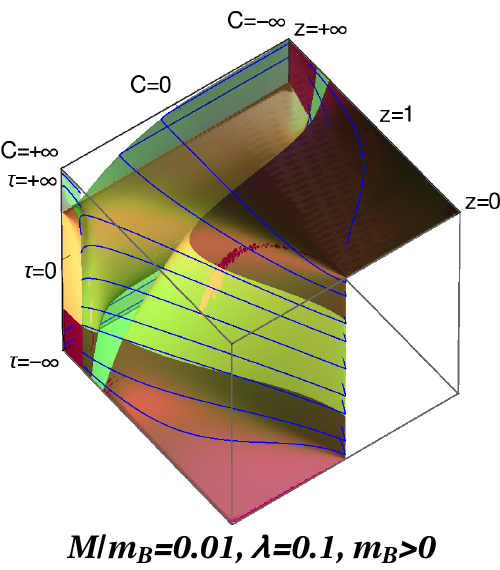}~~~~~~~~~~~~~~~\includegraphics[width=0.3\textwidth]{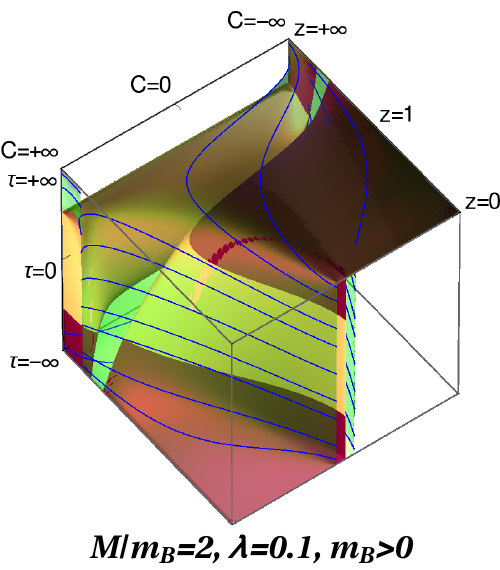}
\\
\includegraphics[width=0.3\textwidth]{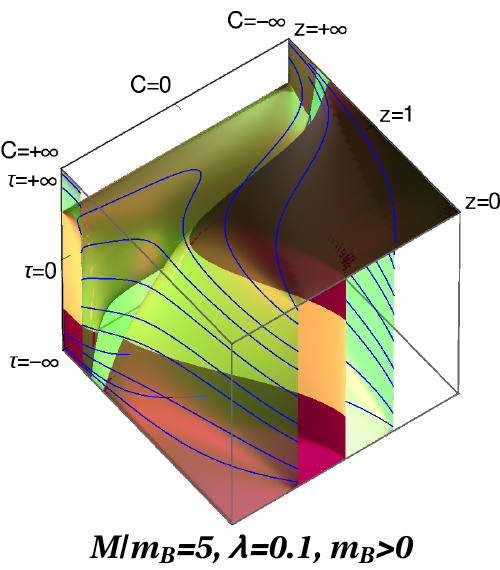}~~~~~~~~~~~~~~~~\includegraphics[width=0.3\textwidth]{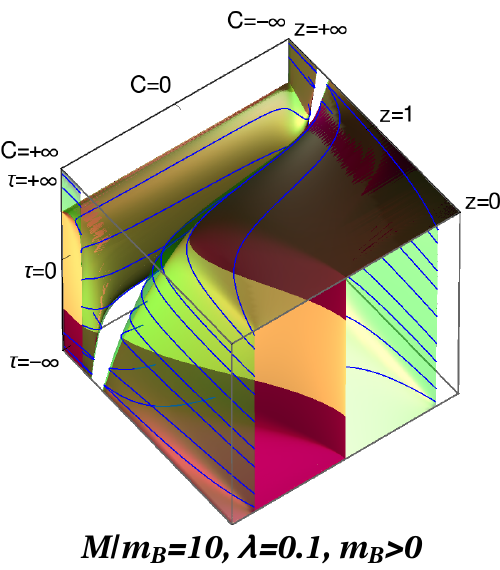}
\caption[On-shell surfaces for $\lambda>0$, $m_\st{B}>0$]{\footnotesize
The surface $\mathscr P (z_a) =0$  of Fig.~\ref{MordorFig2_example}
 for $m_\st{B}>0$, $\lambda = 0.1 >0$ (in yellow/red), together with the on-shell surface (in transparent green, with a few constant-$\tau$ lines in blue), for four choices of the ratio $M_a/m_\st{B}$.}
\label{Mordor_Onshell_Fig_2_example}
\end{figure}

Equations~(\ref{Twinshell_OnshellCondition_dimensionless}) identify each a surface in the 3D space $\left( z_a , C, \tau\right)$.
That same space has `forbidden regions', coinciding with the regions where the polynomial $\mathscr P$, calculated with parameters $z=\frac{\rho_a}{|m_\st{B}|}$, $C=\frac{A_\st{B}}{2m_\st{B}^2}$ and $\tau= |m_\st{B}| \langle p \rangle$, is negative. If the on-shell surface intersected the forbidden region we would be in trouble: it would mean that the reduced phase space has regions where the values of the area of the shell and the other dynamical variables are not acceptable, as the constraint equations~(\ref{ADMconsts}) admit no solution with the boundary conditions set by such a shell.

In Appendix~\ref{MordorAppendix} we plot the on-shell surfaces~(\ref{Twinshell_OnshellCondition_dimensionless}) together with the `forbidden' region $\mathscr P <0$  in the 3D space $\left( z_a , C, \tau\right)$, for all possible choices of dimensionless cosmological constant $\lambda = m_\st{B}^2 \Lambda$, sign of $m_\st{B}$, and value of the rest-mass of the shell $\mu_a = M_a/|m_\st{B}|$. It turns out that \textbf{the on-shell surface never intersects the forbidden region}. This is a very remarkable result, which encourages us to think that the dynamics of our system is well-defined. In Fig.~\ref{Mordor_Onshell_Fig_2_example} we show an example (the same as the right-hand-side of Fig.~\ref{MordorFig2_example}, with $m_\st{B} >0$ and $\lambda = 0.1$) with four choices of rest-mass $M_a$.

What one would like to do now is to solve all equations and identify the minimal core of dynamical variables that are needed for a description of the system, \emph{i.e.}, find the reduced phase space. This cannot be done in isotropic gauge as was done in Sec.~\ref{SubsecSingleShellUniverse} for a single shell. However we can repeat what was done at the end of that section, and concentrate on gauge-independent variables ($\rho_\st{S}$, $\rho_\st{N}$, $A_\st{S}$,  $A_\st{B}$,  $A_\st{N}$) and try to calculate the symplectic form in terms of those variables alone. This turns out to be possible also in the `twin-shell' case.

\subsection{Symplectic form}

The generalization of the pre-symplectic potential~(\ref{IsotropicSymplectic0}) to the case of two shells is:
\begin{equation}
\theta =  -  8\pi \sum_{\beta \in \{ \rm{S},\rm{B},\rm{N}\}} \int_{0}^{\pi}  \Theta_\beta(r)  \left[ \frac 1 3  \mu \,  \sigma \,  \delta \langle p \rangle + \frac{\mu}{\sqrt \sigma} \, \delta A_\beta \right] \d r  \,,
\end{equation}
where
\begin{equation}
\Theta_\beta(r)  = \left\{
\begin{array}{ll}
\Theta(R_\st{S}-r) \,  & \beta = \rm{S} \,,
\\
\Theta(r - R_\st{S})\Theta(R_\st{N}-r) \,  & \beta = \rm{B} \,,
\\
\Theta(r - R_\st{N}) \,  & \beta = \rm{N} \,.
\end{array}
\right.
\end{equation}
We can write
\begin{equation}
\begin{aligned}
\theta 
&=  -  8 \pi    \sum_{\beta \in \{ \rm{S},\rm{B},\rm{N}\}}  \int_{\pi}^{0} \frac{\left(\frac 1 3 \sigma^{3/2}(r) \, \delta \langle p \rangle  + \delta A_\beta  \right) \Theta_\beta(r) \, |\sigma'| \d r}{\sqrt{ A^2_\beta  + \left( {\sfrac 2 3} \langle p \rangle  A_\beta - 8 \, m_\beta  \right) \sigma^{3/2}  + 4 \, \sigma^2 - {\sfrac 1 9}  \left( 12 \, \Lambda - \langle p \rangle^2 \right) \sigma^3 }} 
\\
&=  -  8 \pi  \sum_{\beta \in  \{ \rm{S},\rm{B},\rm{N}\}}   \int_0^\pi \left( \frac{\partial F_\beta [A_\beta,\langle p \rangle ,\sigma]}{\partial \langle p \rangle}  \delta \langle p \rangle  +  \frac{\partial F_\beta [A_\beta,\langle p \rangle ,\sigma]}{\partial A_\beta}  \delta A_\beta \right) | \sigma'| \d r    \,,
\end{aligned}
\end{equation}
where
{\medmuskip=0mu
\thinmuskip=0mu
\thickmuskip=0mu
\begin{equation}
F_\beta[A_\beta ,\langle p \rangle ,\sigma] ~=~   \log \left(\sqrt{ A_\beta^2   + \left( {\sfrac 2 3} \langle p \rangle  A_\beta  - 8 \, m_\beta   \right) \sigma^{3/2}  + 4 \, \sigma^2 - {\sfrac 1 9}  \left( 12 \, \Lambda - \langle p \rangle^2 \right) \sigma^3 }+A_\beta   + {\sfrac 1 3} \langle p \rangle    \sigma^{3/2}\right) \,.
\end{equation}}

The boundary conditions force $\sigma$ to be zero at the poles, and rise monotonically away from the poles up to the location of the two shells, $R_\st{S}$ and $R_\st{N}$. In the `belt' region, $\sigma$ has to be piecewise monotonic except when its value reaches a zero of the polynomial $ A^2_\st{B}  + \left( {\sfrac 2 3} \langle p \rangle  A_\st{B} - 8 \, m_\st{B}  \right) \sigma^{\frac 3 2}  + 4 \, \sigma^2 - {\sfrac 1 9}  \left( 12 \, \Lambda - \langle p \rangle^2 \right) \sigma^3$. A situation of particular interest is when $\Lambda >0$ and $\langle p \rangle^2 < 12 \,\Lambda$, so that there is a maximal positive root of the polynomial whose value is dominated by $\Lambda$ (a cosmological curvature scale). Then a consistent choice is to have $\sigma$  grow monotonically from $R_\st{S}$ to $r_\st{max}$, the location of its absolute maximum, and then decrease monotonically from $r_\st{max}$ to $R_\st{N}$ (see Fig.~\ref{TwinShell_DeterminingIntervalSigma}). This means that our pre-symplectic potential can be written
\begin{equation}
\begin{aligned}
\theta =  -  8 \pi \bigg[  &  \int_0^{\rho^2_\st{S}}\left( \frac{\partial F_\st{S}}{\partial \langle p \rangle}  \delta \langle p \rangle  +  \frac{\partial F_\st{S}}{\partial A_\st{S}}  \delta A_\st{S} \right)  \, d \sigma 
+ \int_{\rho^2_\st{S}}^{\rho^2_\st{max}} \left( \frac{\partial F_\st{B}}{\partial \langle p \rangle}  \delta \langle p \rangle  +  \frac{\partial F_\st{B}}{\partial A_\st{B}}  \delta A_\st{B} \right)  \, d \sigma  +
\\
&
 \int_{\rho^2_\st{N}}^{\rho^2_\st{max}} \left( \frac{\partial F_\st{B}}{\partial \langle p \rangle}  \delta \langle p \rangle  +  \frac{\partial F_\st{B}}{\partial A_\st{B}}  \delta A_\st{B} \right) d \sigma 
+ \int_0^{\rho^2_\st{N}} \left( \frac{\partial F_\st{N}}{\partial \langle p \rangle}  \delta \langle p \rangle  +  \frac{\partial F_\st{N}}{\partial A_\st{N}}  \delta A_\st{N} \right)  d \sigma \bigg]      \,,
\end{aligned}
\end{equation}
and since
\begin{equation}
 \frac{\partial F_\beta [A_\beta,\langle p \rangle , \sigma]}{\partial \langle p \rangle}  = \frac 1 3 \sigma^{3/2} \frac{\partial F_\beta [A_\beta,\langle p \rangle , \sigma]}{\partial A_\beta} \,,
 \end{equation}
 we can write
 \begin{equation}
\begin{aligned}
\theta =  -  8 \pi \bigg[  &  \int_0^{\rho^2_\st{S}} \frac{\partial F_\st{S}}{\partial A_\st{S}}   \left( {\sfrac 1 3} \sigma^{\frac 3 2} \delta \langle p \rangle  + \delta A_\st{S} \right)  \, d \sigma 
+ \int_{\rho^2_\st{S}}^{\rho^2_\st{max}}   \frac{\partial F_\st{B}}{\partial A_\st{B}} \left({\sfrac 1 3} \sigma^{\frac 3 2}  \delta \langle p \rangle  +  \delta A_\st{B} \right)  \, d \sigma  +
\\
&
 \int_{\rho^2_\st{N}}^{\rho^2_\st{max}}  \frac{\partial F_\st{B}}{\partial A_\st{B}}   \left( {\sfrac 1 3} \sigma^{\frac 3 2}  \delta \langle p \rangle  + \delta A_\st{B} \right) d \sigma 
+ \int_0^{\rho^2_\st{N}} \frac{\partial F_\st{N}}{\partial A_\st{N}}  \left( {\sfrac 1 3} \sigma^{\frac 3 2}  \delta \langle p \rangle  +   \delta A_\st{N} \right)  d \sigma \bigg]      \,.
\end{aligned}
\end{equation}
The symplectic form is then
\begin{equation}
\begin{aligned}
\delta \theta =&  -  8 \pi     \left[  \frac{\partial F_\st{S} [\sigma = \rho^2_\st{S}]}{A_\st{S}}  \delta \rho^2_\st{S} \wedge \left( {\sfrac 1 3} \rho^3_\st{S} \delta \langle p \rangle  + \delta A_\st{S} \right)   -   \frac{\partial F_\st{B} [\sigma = \rho^2_\st{S}]}{A_\st{B}}  \delta \rho^2_\st{S} \wedge \left( {\sfrac 1 3} \rho^3_\st{S}   \delta \langle p \rangle  + \delta A_\st{B} \right)   \right] 
\\
&  -  16 \pi \,  \frac{\partial F_\st{B} [\sigma = \rho^2_\st{max} ]}
{A_\st{B}} \delta \rho^2_\st{max} \wedge  \left( {\sfrac 1 3} \rho_\st{max}^3 \delta \langle p \rangle  + \delta A_\st{B} \right) 
\\
& -  8 \pi  \left[ \frac{\partial F_\st{N} [\sigma = \rho^2_\st{N}]}{A_\st{N}}  \delta \rho^2_\st{N} \wedge  \left( {\sfrac 1 3} \rho^3_\st{N}  \delta \langle p \rangle  + \delta A_\st{N} \right)  - \frac{\partial F_\st{B} [\sigma = \rho^2_\st{N}]}{A_\st{B}} \delta \rho^2_\st{N} \wedge   \left( {\sfrac 1 3} \rho^3_\st{N} \delta \langle p \rangle  + \delta A_\st{B} \right)  \right]    \,.
\end{aligned}
\end{equation}
We can prove that $\rho_\st{max}^2$ completely disappears from the symplectic form. In fact $\rho_\st{max}^2$ is a solution of the equation
\begin{equation}
A^2_\st{B}  + \left( {\sfrac 2 3} \langle p \rangle  A_\st{B} - 8 \, m_\st{B}  \right) \rho_\st{max}^3  + 4 \, \rho_\st{max}^4- {\sfrac 1 9}  \left( 12 \, \Lambda - \langle p \rangle^2 \right) \rho_\st{max}^6 = 0 \,.
\end{equation}
Varying the above equation w.r.t. $\rho_\st{max}$, $A_\st{B}$ and $\langle p \rangle$ we get an identity for $\delta \rho_\st{max}$:
\begin{equation}
\delta \rho_\st{max}^2 = f[\rho_\st{max},A_\st{B},\langle p \rangle] \left( {\sfrac 1 3} \rho_\st{max}^3 \delta \langle p \rangle + \delta A_\st{B} \right) \,,
\end{equation}
and therefore the only term containing $\rho_\st{max}$ vanishes:
\begin{equation}
\begin{aligned}
 &-  16 \pi \,  \frac{\partial F_\st{B} [\sigma = \rho^2_\st{max} ]}
{A_\st{B}} \delta \rho^2_\st{max} \wedge  \left( {\sfrac 1 3} \rho_\st{max}^3 \delta \langle p \rangle  + \delta A_\st{B} \right) 
\\
&= -  16 \pi \,  \frac{\partial F_\st{B} [\sigma = \rho^2_\st{max} ]}
{A_\st{B}} f[\rho_\st{max},A_\st{B},\langle p \rangle] \left( {\sfrac 1 3} \rho_\st{max}^3 \delta \langle p \rangle + \delta A_\st{B} \right)\wedge  \left( {\sfrac 1 3} \rho_\st{max}^3 \delta \langle p \rangle  + \delta A_\st{B} \right) =0  \,.
\end{aligned}
\end{equation}
We can finally use the boundary conditions at the poles, $A_\st{S} = m_\st{S} = A_\st{N} = m_\st{N} =0$, and we get the following nondegenerate 2-form ($\omega = \delta \theta$):
%
{\medmuskip=0mu
\thinmuskip=0mu
\thickmuskip=0mu
\begin{equation}\label{SymplecticFormTwinShell}
\begin{aligned}
\omega  =  -  \frac{8 \pi}{3}     \bigg[  & \frac{ \rho^3_\st{S} \delta \rho^2_\st{S} \wedge \delta \langle p \rangle }{\sqrt{  4 \, \rho_\st{S}^4 - {\sfrac 1 9}  \left( 12 \, \Lambda - \langle p \rangle^2 \right) \rho_\st{S}^6 }}   ~-~   \frac{   \rho^3_\st{S}   \delta \rho^2_\st{S} \wedge  \delta \langle p \rangle + 3 \, \delta \rho^2_\st{S} \wedge \delta A_\st{B}}{\sqrt{ A^2_\st{B}  + \left( {\sfrac 2 3} \langle p \rangle  A_\st{B} - 8 \, m_\st{B}  \right) \rho_\st{S}^3 + 4 \, \rho_\st{S}^4 - {\sfrac 1 9}  \left( 12 \, \Lambda - \langle p \rangle^2 \right) \rho_\st{S}^6 }}   +
\\
& \frac{  \rho^3_\st{N} \delta \rho^2_\st{N} \wedge \delta \langle p \rangle }{\sqrt{  4 \, \rho_\st{N}^4 - {\sfrac 1 9}  \left( 12 \, \Lambda - \langle p \rangle^2 \right) \rho_\st{N}^6 }}  ~-~  \frac{   \rho^3_\st{N}   \delta \rho^2_\st{N} \wedge  \delta \langle p \rangle + 3 \, \delta \rho^2_\st{N} \wedge \delta A_\st{B}}{\sqrt{ A^2_\st{B}  + \left( {\sfrac 2 3} \langle p \rangle  A_\st{B} - 8 \, m_\st{B}  \right) \rho_\st{N}^3 + 4 \, \rho_\st{N}^4 - {\sfrac 1 9}  \left( 12 \, \Lambda - \langle p \rangle^2 \right) \rho_\st{N}^6 }} \bigg]  \,.
\end{aligned}
\end{equation}}
The above 2-form is nondegenerate in the 4-dimensional phase space coordinatized by $\langle p \rangle$, $A_\st{B}$, $\rho_\st{S}$ and $\rho_\st{N}$. To reach the above expression we used every constraint that was at our disposal (the solution of the Hamiltonian, diffeomorphism and conformal constraint, and the two diffeomorphism jump conditions), except the two jump conditions associated to the Hamiltonian constraint. Notice that we didn't need to use a diffeomorphism gauge fixing to get a nondegenerate symplectic form, because we were able to recast the pre-symplectic form in a reparametrization-invariant form. In other terms, we avoided having to completely gauge fix our constraints by expressing the symplectic form in terms of a maximal system of gauge-invariant quantities.

\section{Breakdown of the ADM description}\label{SecTheProblem}

\begin{figure}[b!]\center
~~~~~~~~~~~~~\includegraphics[width=\textwidth]{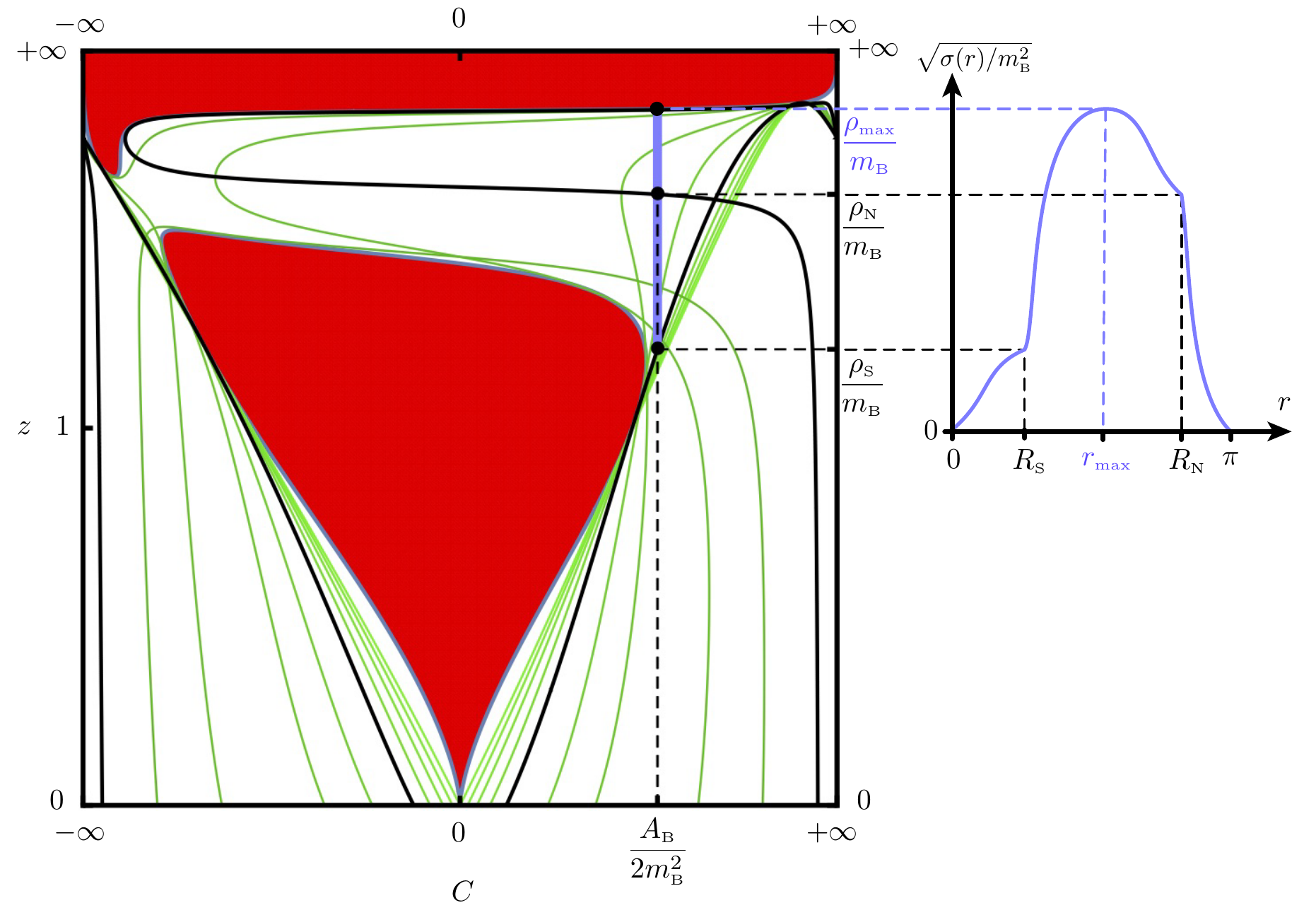}
\caption[Determining the codomain of $\sigma$ in the twin-shell universe]{\footnotesize
A plot of the on-shell curves at a fixed York time $\tau =  0.46$ for a set of values of the rest mass $M_a/m_\st{B}$ (between 0 and 20), and for  $\lambda = 0.1$, $m_\st{B}>0$.
The excluded region $\mathscr P <0$ is in red.
Given the values of the rest masses of the two shells (in the figure $M_\st{S} = 2.5 \, m_\st{B}$ and  $M_\st{N} = 20 \, m_\st{B}$), specifying the value of the integration constant $A_\st{B}$ in the belt completely fixes  $\rho_\st{S}$, $\rho_\st{N}$ and $\rho_\st{max}$. Then the interval of values of the areal radius coordinate $\sigma(r)$ of the metric in the belt is fixed (light-blue strip). 
$\sigma$ will go from $\rho_\st{N}$ to a maximum given by the border of the excluded region (where $\sigma'$ is allowed to vanish), and then will go down until it reaches $\rho_\st{S}$. A  choice of $\sqrt{\sigma(r)}$ compatible with the boundary conditions imposed by the values of $\rho_\st{S}$, $\rho_\st{N}$ and $\rho_\st{max}$ is showed on the right.
}\label{TwinShell_DeterminingIntervalSigma}
\end{figure}

In this section I will discuss the conditions under which the ADM description of the system breaks down. To do this, I need first to show how the on-shell relations~(\ref{Twinshell_OnshellCondition_dimensionless}) are to be used to provide boundary conditions for the metric in a context with two shells. 
Consider a constant-York-time  slice $\tau = \text{\it const.}$. In Fig.~\ref{TwinShell_DeterminingIntervalSigma} I plot the `forbidden' region $\mathscr P <0$ in red in the plane $(C,z)$. In green I show a series of on-shell curves, solutions of~(\ref{Twinshell_OnshellCondition_dimensionless}) for different values of the rest-mass (normalized by $|m_\st{B}|$): $M/|m_\st{B}|$. Among these, two curves will correspond to the rest mass of the two shells, $M_\st{N}$ and $M_\st{S}$. I plot those in black. If we choose a value of $A_\st{B}$, through Eq.~(\ref{Twinshell_OnshellCondition_dimensionless}) we are also fixing the value of the areal radii of the two shells, $\rho_\st{S}$ and $\rho_\st{N}$, which can be read in the diagram as the ordinates of the corresponding points on the two black on-shell curves. This is like fixing the total energy of a one-dimensional system; the relation between position (areal radius) and momentum (given by $A$) is thereafter completely determined. The constraints of the system do not allow for independent behaviour of the two shells: they are `interlocked'.
Moreover, if $\lambda >0$, we also fix a \emph{maximum} areal radius $\rho_\st{max}$ that the metric can support, which is essentially determined by the cosmological constant (in Fig.~\ref{TwinShell_DeterminingIntervalSigma} it is the border of the top disconnected component of the red forbidden region).  Given all this data, we can determine an attainable form for the $\theta\theta$ component of the metric ($\sigma(r)$, the areal radius squared): it will monotonically interpolate $\sigma = 0$ with $\sigma = \rho_\st{S}^2$ (resp. $\rho_\st{N}^2$) from $r =0$ (resp. $\pi$) to $r= R_\st{S}$ (resp. $R_\st{N}$). Then its derivative will have, at $r= R_\st{S}$  (resp. $r= R_\st{N}$) a certain jump determined by Eqs.~(\ref{EqsForGammasAndKappas}). In the region in between (the `belt' region) $\sigma$ will go from $\sigma(R_\st{S}) = \rho^2_\st{S}$ to a maximum $\sigma(r_\st{max}) = \rho^2_\st{max}$ and then down to $\sigma(R_\st{N}) = \rho^2_\st{N}$. Away from $r=r_\st{max}$, $\sigma$ will be monotonic. All of this is  illustrated by the Cartesian diagram on the right of Fig.~\ref{TwinShell_DeterminingIntervalSigma}.
Notice that, while interpolating in the belt region from one shell to the other, we could  have alternatively avoided having the areal radius reach the maximum value $\rho_\st{max}^2$ and bounce back. This is an acceptable choice if $\rho_\st{N} \neq \rho_\st{S}$, because the areal radius could monotonically interpolate between $\rho_\st{S}$ and $\rho_\st{N}$. But as we can see in the diagram above, the two black on-shell curves intersect at a point, which means that there exists a value of $A_\st{B}$ such that $\rho_\st{N} = \rho_\st{S}$ even though $M_\st{N} \neq M_\st{S}$. Then in this case we are forced to have the areal radius grow up to  $\rho_\st{max}^2$ and back, otherwise it could not possibly be interpolating between $\rho_\st{N}$ and $\rho_\st{S}$ while being monotonic. I conclude that the only consistent choice is that $\sigma$ \emph{always} bounces off the  value $\rho_\st{max}^2$, even when $\rho_\st{N} \neq \rho_\st{S}$.

\begin{figure}[b!]\center
~~~~~~~~~~~~~\includegraphics[width=\textwidth]{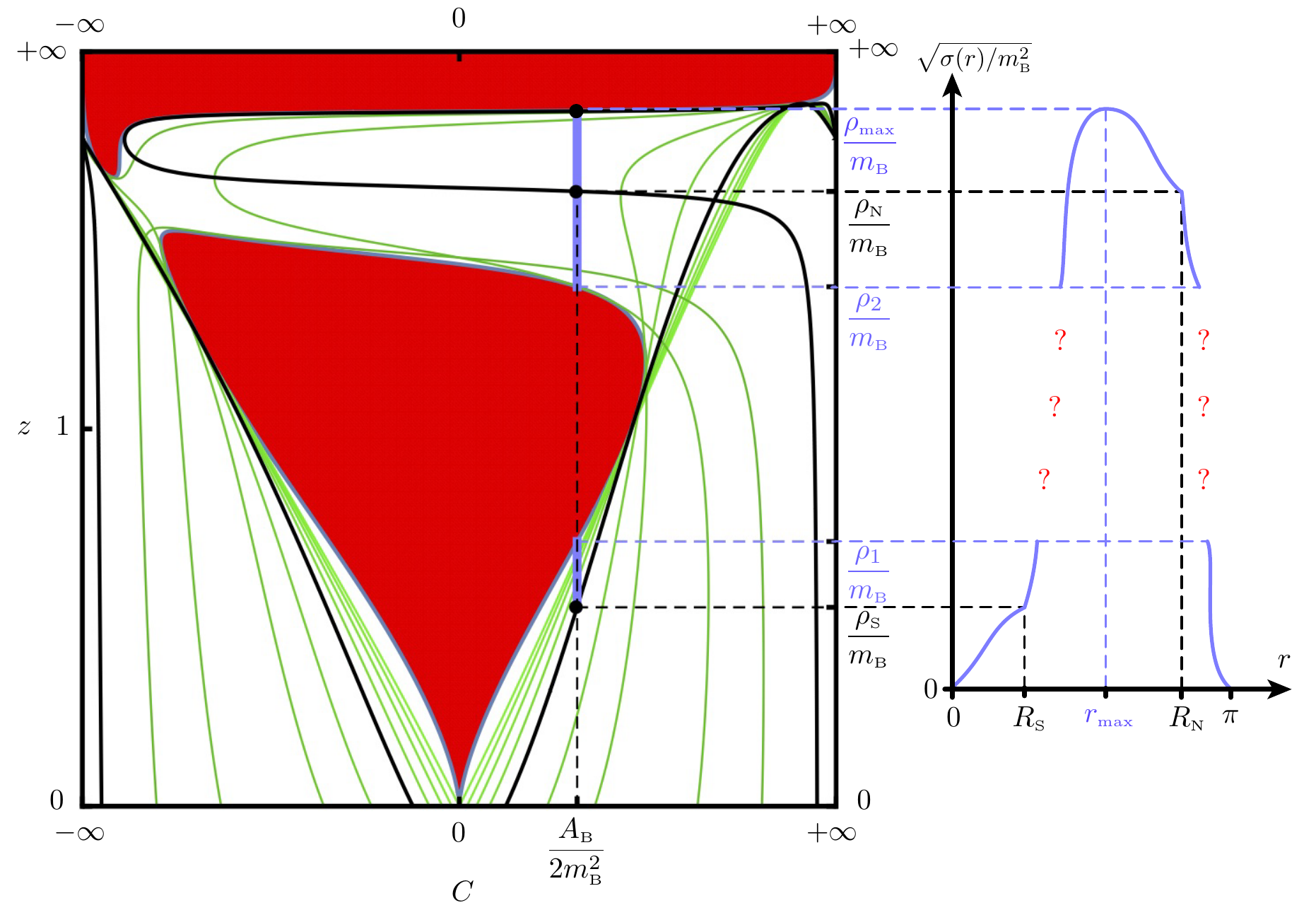}
\caption[Determining the codomain of $\sigma$ in the twin-shell universe]{\footnotesize
Same plot as above, but with a value of $A_\st{B}$ such that the $C = \frac{A_\st{B}}{2 m_\st{B}}$ line crosses the bottom forbidden region (in red). The two points at which this crossing happens have $z = \frac{\rho_1}{m_\st{B}}$ and $z = \frac{\rho_2}{m_\st{B}}$.}\label{ProblemFig2}
\end{figure}

I am now ready to present the issue. Consider the diagram of Fig.~\ref{ProblemFig2}.
Now the chosen value of $A_\st{B}$ is such that the forbidden region crosses the line that connects $\rho_\st{S}$ with $\rho_\st{N}$. In this situation there is no acceptable solution to the constraint equations! In fact, the areal radius of the metric $\sqrt{\sigma}$ cannot  take all the values that are included in the interval $\left( \rho_\st{S},\rho_\st{N} \right)$, because a section of this interval is excluded.
There exists no metric that solves the Lichnerowicz--York equation in this situation. 
In the shape-dynamical interpretation of this system the spatial metric  is not itself physical, only its shape degrees of freedom are, and they live in a reduced shape-phase space, which is represented by the green on-shell surface, which never crosses the forbidden region and seems to be globally well-defined. On the other hand, in the ADM interpretation the spatial metric is the pull-back of the spacetime metric on a CMC hypersurface, and the fact that it is not well-defined implies that there is no spacetime metric, and the solution is not an acceptable solution of Einstein's equation. We identified a new point of departure between Shape Dynamics and GR: when the dynamical solution enters this region where the areal radius should interpolate between values that surround the forbidden region, the SD description is well-defined, while the GR one is not.

Notice that, as can be seen from the diagrams in Appendix~\ref{AppendixOnshellSurfaces}, the only case in which this departure is possible is that with positive Misner--Sharp mass $m_\st{B} >0$ and positive but small cosmological constant $\lambda >0$, $\lambda \ll  1$. The other choices of $m_\st{B}$ and $\lambda$ do not give rise to a `concave' allowed region where the on-shell surfaces of the two shells are separated by the forbidden region. Interestingly, the $m_\st{B} >0$, $0<\lambda \ll 1$ case is particularly physically relevant, as it seems to match our universe more closely.

\section{Outlook and conclusions}

In this article I presented my most advanced understanding of gravitational collapse in a relational, compact universe.  The simplest nontrivial compact spherically symmetric model  has the topology of a 3-sphere, has a positive cosmological constant and contains one spherical  thin shells of dust. The compact boundary conditions in this case are too restrictive, and the momentum of the shell is constrained to be zero, so that the system ends up deprived of dynamical degrees of freedom. Nevertheless, I find a family of dynamically trivial solutions  which are parametrized by the rest mass of the shell, and I also find a bound on this rest mass that forces the associated length scale to be smaller than the cosmological horizon scale associated to $\Lambda$. To my knowledge, such a spatially compact solution of General Relativity with one spherical shell of dust has not been studied before.

I then move on to study a slightly less trivial system, which has enough dynamical degrees of freedom that it can model gravitational collapse. This model involves two thin spherical shells and, again, a positive cosmological constant. It turns out that such a model contains the bare minimum structure that is necessary to model gravitational collapse in a closed universe: if one of the shells has a rest mass that is much larger than that of the other, this shell will play the role of `spectator' (\emph{i.e.} `fixed stars' or `rest of the universe'), while the `light' shell will be able to undergo collapse. In such a situation it becomes meaningful to say that one shell collapsed, because the `heavy' shell provides a reference scale.

I was able to study the reduced phase space of such a system, characterizing it in a geometric way as a couple of surfaces (one for each shell) in the 3D space $A_\st{B}$ (related to the momentum of both shells), $\rho_a$ (the areal radius of the shells) and $\langle p \rangle$ (the York time). The shape of the surface associated to each shell depends on the rest mass of that shell. The two shells share the same value of $A_\st{B}$, but that value corresponds to different $\rho_a$'s, depending on their rest mass.  The 3D ambient space has some `forbidden' regions, whose shape depends on the sign of the cosmological constant and of the Misner--Sharp mass of the system. In those regions, there can be no metric which is a solution of the ADM constraints and has the prescribed value of the integration constant $A_\st{B}$ while reaching the corresponding value of areal radius $\sigma(R_a) =\rho_a$. Fortunately, the on-shell surfaces that describe reduced phase space never cross those regions. This  is a consistency check for the system. There is, however, an issue with the ADM description. Even though no on-shell surface crosses the forbidden region, the values of the rest masses of the two shells can be such that the corresponding on-shell surfaces `surround' the forbidden region. More precisely, the constant-$A$ line connecting the point on one surface with the point on the other surface intersects the forbidden region. This means that, even though the areal radius of the metric $\sigma$ can take the values $\sigma(R_a)=\rho_a$ demanded by the boundary conditions, in order to interpolate between these two values it would have to go through forbidden values.
In other words, the combination of values of $A_\st{B}$ and $\rho_a$ is such that there is no solution of the ADM constraints which is compatible with the boundary conditions imposed by the size and  momenta of the shells. This should not be, in principle, a problem for Shape Dynamics, which does not rely on the ADM constraints holding at all times.

%
%
%
%

\section*{Acknowledgments}

A big thank to Henrique Gomes for support, discussion and active involvement in the development of this project during all its phases. Thanks to Tim Koslowski, Sean Gryb and David Sloan for discussions and their encouragement. This research was supported in part by Perimeter Institute for Theoretical Physics. Research at Perimeter Institute is supported by the Government of Canada through the Department of Innovation, Science and Economic Development Canada and by the Province of Ontario through the Ministry of Research, Innovation and Science.
Support was partially granted also from a Marie Curie fellowship of the Istituto Nazionale di Alta Matematica.

\providecommand{\href}[2]{#2}\begingroup\raggedright\endgroup

\newpage

\appendix

\section{Appendix: boundary conditions at the poles}\label{AppendixBoundaryConditions}

In this Appendix we will present all the evidence we collected so far in favour of the boundary conditions $A_\st{N} = A_\st{S} = m_\st{N} = m_\st{S} =0$, and provide a further argument which, we believe, closes the issue.

\begin{figure}[ht!]
\begin{minipage}{\textwidth}\center
$\bm \lambda < 0 \, , ~ \bm m>0$
\\
\includegraphics[width=0.35\textwidth]{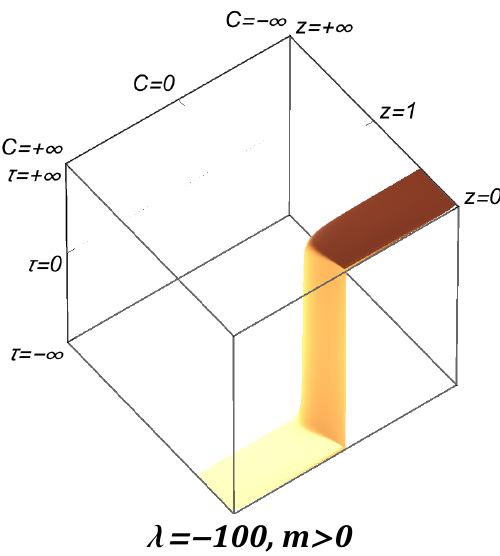}~~~~~~~~~~~~~~~~
\includegraphics[width=0.35\textwidth]{Mordor2.pdf}
\end{minipage}
\caption[The surface $\mathscr P (z) =0$  for $m>0$, $\lambda < 0$]{\footnotesize
The surface $\mathscr P (z) =0$ for \textbf{positive Misner--Sharp mass}  and \textbf{negative cosmological constant.}}\label{MordorFig1}
\end{figure}

\begin{figure}[ht!]
\begin{minipage}{\textwidth}\center
$\bm \lambda \geq 0 \, , ~ \bm m>0$
\\
\includegraphics[width=0.35\textwidth]{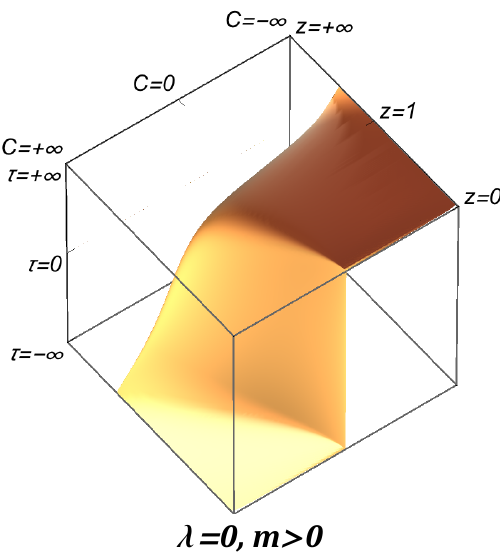}~~~~~~~~~~~~~~~~
\includegraphics[width=0.35\textwidth]{Mordor4.pdf}
\\
\includegraphics[width=0.35\textwidth]{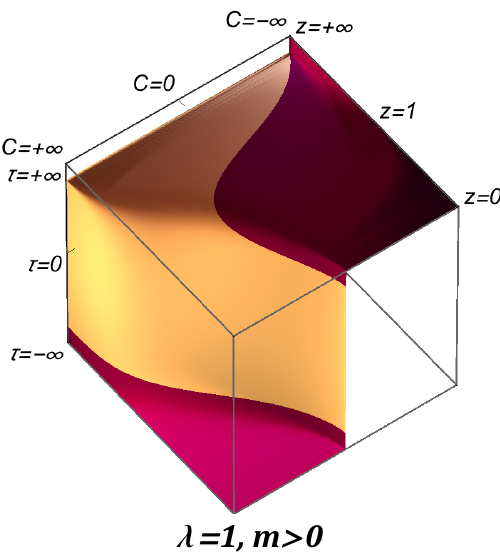}~~~~~~~~~~~~~~~~
\includegraphics[width=0.35\textwidth]{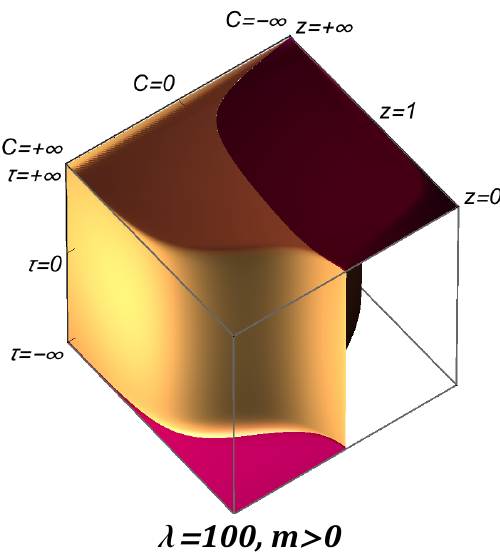}
\end{minipage}
\caption[The surface $\mathscr P (z) =0$  for $m>0$, $\lambda \geq 0$]{\footnotesize
The surface $\mathscr P (z) =0$ for \textbf{positive Misner--Sharp mass}  and four choices of \textbf{zero or positive cosmological constant.}  The part of the surface where $\tau^2 < 12 \Lambda$ is in yellow, while the part $\tau^2 > 12 \Lambda$ is in red.}
\label{MordorFig2}
\end{figure}

\begin{figure}[ht!]
\begin{minipage}{\textwidth}\center
$\bm \lambda > 0 \, , ~ \bm m<0$
\\
\includegraphics[width=0.35\textwidth]{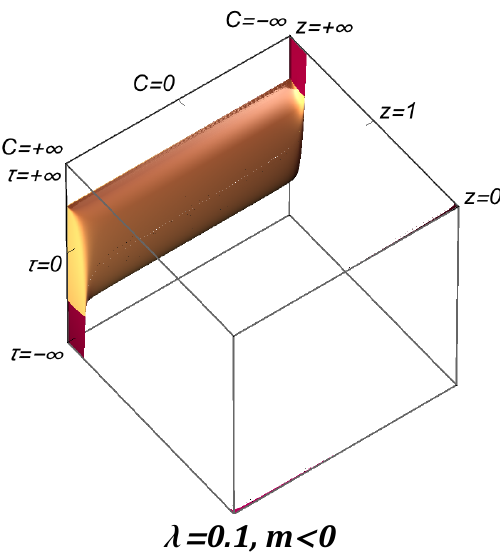}~~~~~~~~~~~~~
\includegraphics[width=0.35\textwidth]{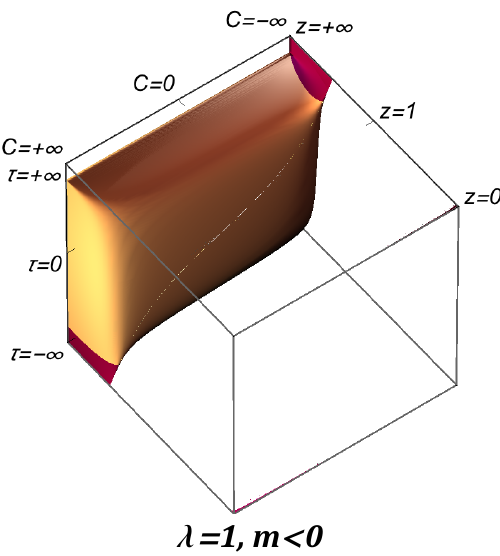}\\
\includegraphics[width=0.35\textwidth]{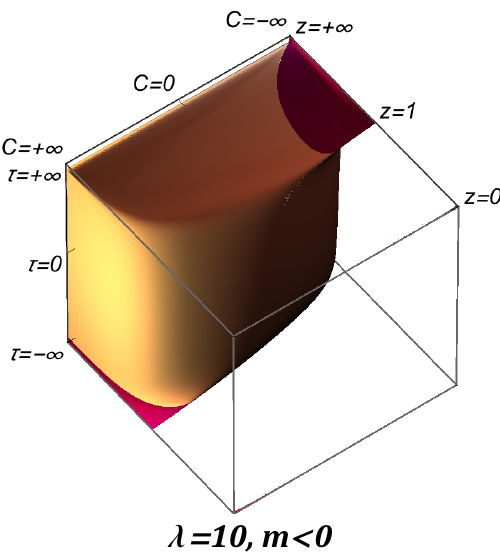}~~~~~~~~~~~~~
\includegraphics[width=0.35\textwidth]{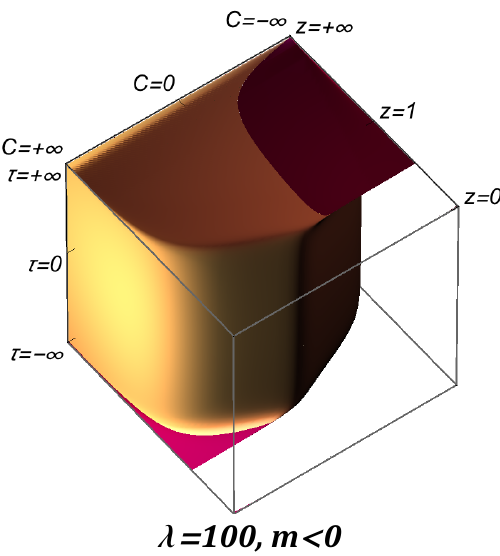}
\end{minipage}
\caption[The surface $\mathscr P (z) =0$  for $m<0$, $\lambda > 0$]{\footnotesize
The surface $\mathscr P (z) =0$ for \textbf{negative Misner--Sharp mass}  and 4 choices of \textbf{positive cosmological constant.}  The negative or zero cosmological constant cases are not included because the $\mathscr P >0$ identically in those cases.}\label{MordorFig3}
\end{figure}

\begin{figure}[ht!]
\begin{center}
\includegraphics[width=0.35\textwidth]{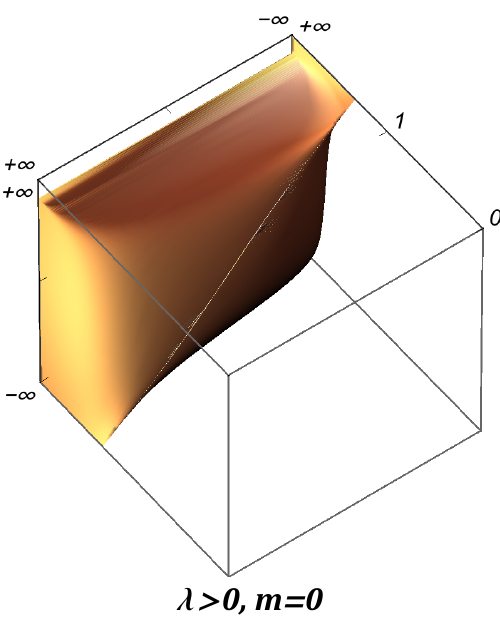}
\end{center}
\caption[The surface $\mathscr P (z) =0$  for $m=0$, $\lambda > 0$]{\footnotesize
The surface $\mathscr P (z) =0$ for \textbf{zero Misner--Sharp mass}  and \textbf{positive cosmological constant}.
In this case we used $|\Lambda|$ to make all variables dimensionless:  $z = \sqrt{|\Lambda|} \sqrt{\sigma}$, $C = |\Lambda| \, A /2$, $\tau = \langle p \rangle/\sqrt{|\Lambda|}$. Only the positive-$\lambda$ case is interesting, because if $\Lambda<0$ the polynomial is always positive. Similarly, the case $\Lambda = m = 0$ is trivial because in this case $\mathcal P$ is identically positive.
}\label{MordorFig4}
\end{figure}

The general solution~(\ref{SolConstraints}) to the ADM constraints involves a metric that takes the  form
\begin{equation}\label{MetricSolutionOfADMConstraints}
\d s^2 =  \frac{\sigma (\sigma')^2 \d r^2}{A^2  + \left( {\sfrac 2 3} \langle p \rangle  A - 8 \, m  \right) \sigma^{\frac 3 2}  + 4 \, \sigma^2 - {\sfrac 1 9}  \left( 12 \, \Lambda - \langle p \rangle^2 \right) \sigma^3 } 
+ \sigma \left( \d \theta^2 + \sin^2 \theta \d \phi^2 \right) \,.
\end{equation}
The component $\sigma$ is allowed to go to zero at the poles $r=0,\pi$ only if the polynomial~(\ref{MordorPoly}), and with it the denominator of the $\d r^2$ term, is positive around $\sigma =0$. Looking at the regions of positivity of $\mathscr P(\sqrt{\sigma}/m)$ in Fig.~\ref{MordorFig1}--\ref{MordorFig4} we see that on the plane $\sigma = 0$ the polynomial is always positive, unless $A=0$, in which case it is zero. It is easy to see that the on-shell curves which end at $A=\sigma=0$ will do so in such a way that the polynomial will stay positive all the time. If $m>0$, this means that the behaviour of $\sigma$ for small $A$'s will have to be $\sigma \xrightarrow[A\to 0]{} \left( \frac{\beta}{8m}\right)^{\frac 2 3} |A|^{\frac 4 3} + \mathcal O (|A|^{{\frac 2 3} +\epsilon})$, where $0 \leq \beta < 1$ (while if $m \leq 0$ there is no constraint on the asymptotics of $\sigma$).

For small values of the areal radius (near the poles), we can ignore the term $- {\sfrac 1 9}  \left( 12 \, \Lambda - \langle p \rangle^2 \right) \sigma^3$ in~(\ref{MetricSolutionOfADMConstraints}), and the three independent curvature invariant \emph{densities} take the form:
\begin{equation}
\begin{aligned}
R_1 = \sqrt g  \, R  &= \frac{\sin \theta \, |\sigma'|}{2 \, \sigma^{3/2}} \frac{3 \, A^2 }{\sqrt{ A^2+ {\sfrac 2 3} B \sigma^{3/2}+4 \sigma^2}}  \,,
\\
R_2 = \sqrt g \,R^i{}_j R^j{}_i &= \frac{\sin \theta \, |\sigma'|}{8 \, \sigma ^{9/2}}  \frac{27 A^4+6 A^2 B \sigma ^{3/2} +B^2 \sigma ^3}{\sqrt{9 A^2+\sigma ^{3/2} (6 A \langle p\rangle -72 m)+36 \sigma ^2}} \,,
\\
R_3 = \sqrt g \, R^i{}_j R^j{}_k  R^k{}_i  &= \frac{\sin \theta \, |\sigma'|}{96 \, \sigma ^{15/2}} \frac{297 A^6+135 A^4 B \sigma ^{3/2}+27 A^2 B^2 \sigma ^3 + B^3 \sigma ^{9/2} }{\sqrt{9 A^2 + 6 B \sigma ^{3/2} +36 \sigma ^2}} \,,
\end{aligned}
\end{equation}
where $B=\left( A \langle p \rangle - 12 m\right)$. If $A \neq 0$, all these quantities diverge as $\sigma \to 0$ and we have a curvature singularity at the the poles.\footnote{It is not hard to convince oneself that there is no way to have the $|\sigma'|$ term at the numerator cancel the divergence of the denominator while $\sigma \to 0$. In fact if $\sigma \sim r^n$, then $|\sigma'|/\sigma^{3/2}$ is finite if $n \leq -2$, but then $\sigma$ diverges as $r\to 0$.}
If $A=0$ and $m <0$ the first curvature invariant is zero, but the other two are still divergent. If $m>0$ and $\sigma \sim \left(  \beta/8m\right)^{\frac 2 3} |A|^{\frac 4 3}+ \mathcal O (|A|^{{\frac 4 3} +\epsilon})$ the three terms diverge like
\begin{equation}
R_1   \sim \frac{12 m}{A \sqrt{1-\beta } \beta } \,, ~~
R_2 \sim \frac{48 \left(12 -4 \beta +\beta ^2\right) m^3}{A^3 \sqrt{1-\beta } \beta ^3}  \,, ~~
R_3  \sim \frac{384 \left(
88 - 60 \beta - 18 \beta^2 + \beta^3
\right) m^5}{A^5 \sqrt{1-\beta } \beta ^5} \,.
\end{equation}
So the metric (\ref{MetricSolutionOfADMConstraints}) always has a curvature singularity at the poles, for any value of  the parameters $A$  and $m$, unless $A=m=0$. This should be a sufficient reason to take $A=m=0$ as our boundary conditions around the poles, however Shape Dynamics is concerned with the conformal geometry of the metric, and this is regular (conformally flat) even in presence of a curvature singularity. From the perspective of conformal geometry, what the curvature singularity does is to make the theory lose predictivity: in fact the value of $A = A(t)$ at the poles is not fixed by any dynamical equation, and needs to be specified by hand.

To better understand this loss of predictivity, turn now to the vacuum diffeomorphism constraint, $\nabla_j p^j{}_i =0$.  The solution~(\ref{SolConstraints}) of this constraint is:
\begin{equation}
p^j{}_i = \mu \left[ \left( {\sfrac 1 3} \langle p \rangle \, \sigma  + \frac{A}{\sqrt \sigma} \right)  \, \delta^j_r \delta^r_i +    \left( {\sfrac 1 3} \langle p \rangle \, \sigma  - {\sfrac 1 2}  \frac{A}{\sqrt \sigma}  \right)   \left( \delta^j_\theta \delta^\theta_i  + \delta^j_\phi \delta^\phi_i \right) \right] \sin \theta  \,.
\end{equation}
There is one spherically-symmetric ($X^i = \delta^i{}_r X(r)$) conformal killing vector of the $S^3$ metric:
\begin{equation}
\nabla^i X^j + \nabla^j X^i - {\sfrac 2 3} g^{ij} \nabla_k X^k =0  \qquad \Rightarrow \qquad
X^i = c \, \frac{\sqrt{\sigma}}{\mu} \,  \delta^i{}_r \,,
\end{equation}
(in isotropic gauge this is just $X^i = c \, \sin r \,  \delta^i{}_r$). This vector field is well-behaved at the poles, where $\sigma \to 0$. Now take the vector field $Y^i = p^i_j X^j$. Its coordinate expression is
\begin{equation}
Y^i =  c \, \delta^i{}_r \left( {\sfrac 1 3} \langle p \rangle \, \sigma^{\frac 3 2}  + A  \right)   \sin \theta  \,.
\end{equation}
The divergence of $Y^i$ is
\begin{equation}
\nabla_i Y^i = (\nabla_i p^i_j) X^j + p^{ij} \nabla_i X_j =  (\nabla_i p^i_j) X^j + {\sfrac 1 3} p \nabla_k X^k 
=  (\nabla_i p^i_j) X^j + {\sfrac 1 3} \langle p \rangle  \nabla_k X^k  \, \sqrt g \,,
\end{equation}
integrating over a spherical region centred around the origin:
{\medmuskip=0mu
\thinmuskip=0mu
\thickmuskip=0mu
\begin{equation}
\begin{aligned}
&\int_{r\leq R} \nabla_i Y^i \d^3 x ~ = \int_{r\leq R}  \left[  (\nabla_i p^i_j) X^j + {\sfrac 1 3} \langle p \rangle    \nabla_k X^k \right] \d^3 x ~ = \int_{r\leq R}  (\nabla_i p^i_j) X^j \d^3 x + c {\sfrac {4\pi} 3} \langle p \rangle \, \sigma^{\frac 3 2} (R)
\\
& \qquad \qquad \shortparallel
\\
&\int_{r= R}  Y^i \d \Sigma_i ~=~
4 \pi c \,\left( {\sfrac 1 3} \langle p \rangle \, \sigma^{\frac 3 2} (R)  + A  \right)  
\end{aligned}
\end{equation}}
we conclude that 
\begin{equation}
\int_{r\leq R}  (\nabla_i p^i_j) X^j \d^3 x  = 4 \pi c \, A \,.
\end{equation}
Now notice that, if the region of integration was the annular region $R_1 \leq r \leq R_2$, the result would have been
\begin{equation}
\int_{R_1 \leq r \leq R_2}  (\nabla_i p^i_j)  X^j \d^3 x  = 0\,.
\end{equation}
The same holds for any region $\Omega$ which does not include the pole. We conclude that 
\begin{equation}\label{DiffeoConstraintWithDiracDeltaSource}
 (\nabla_i p^i_j)  X^j  = 4 \pi c \, A \, \delta^{(3)} (\vec r) \,.
\end{equation}
This result is analogue to what one gets when considering the vacuum Poisson equation on $\mathbbm{R}^3$ in polar coordinates:
\begin{equation}
\Delta V = \frac{\partial^2 V}{\partial x^2} + \frac{\partial^2 V}{\partial y^2} +\frac{\partial^2 V}{\partial z^2} = {\frac {1}{r^{2}}}{\frac {\partial }{\partial r}}\left(r^{2}{\frac {\partial V}{\partial r}}\right)+{\frac {1}{r^{2}\sin \theta }}{\frac {\partial }{\partial \theta }}\left(\sin \theta {\frac {\partial V}{\partial \theta }}\right)+{\frac {1}{r^{2}\sin ^{2}\theta }}{\frac {\partial ^{2} V}{\partial \varphi ^{2}}} = 0 \,,
\end{equation}
if $V$ is spherically symmetric, the equation reduces to ${\frac {1}{r^{2}}}{\frac {\partial }{\partial r}}\left(r^{2}{\frac {\partial V}{\partial r}}\right)=0$, which admits the general solution:
\begin{equation}
V = \frac{c_1}{r} + c_2 \,.
\end{equation}
This solution has two integration constants, but they can both be fixed by appropriate boundary conditions: $V \xrightarrow[r\to \infty]{} 0$ implies $c_2 =0$ and regularity at the origin implies $c_1 =0$. If we insist on having $c_1 \neq 0$, we find out that we are not solving the original equation (in vacuum), but an equation with some sources concentrated at the origin:
\begin{equation}
\Delta V = -4 \pi \, c_1 \, \delta^{(3)} (\vec r) \,,
\end{equation}
in fact, using cartesian coordinates:
\begin{equation}
\Delta \left(\frac{c_1}{r} + c_2  \right) = - c_1 \, \vec \partial \cdot \left(\frac{x}{r^3},\frac{y}{r^3},\frac{z}{r^3} \right) 
\end{equation}
and integrating over a sphere of radius $R$:
\begin{equation}
-c_1 \int_{r\leq R}   \vec \partial \cdot  \left(\frac{x}{r^3},\frac{y}{r^3},\frac{z}{r^3} \right) d^3 x = - \frac{c_1}{r^2} \int_{r=R} d\Sigma = - 4 \pi \, c_1 \,.
\end{equation}
The reason for this is the fact that the spherical coordinate patch covers all of $\mathbbm R^3$ except the origin, which lies on the border of the coordinate chart. Then the elliptic equation $\nabla V =0$ turns into a boundary-value problem, depending on the boundary conditions we choose to impose at $r=0$ and $r=\infty$. If we choose $c_1 \neq 0$, we have effectively changed the vacuum equation into one with a Dirac-delta source concentrated at the origin. Such an equation still coincides with the vacuum Poisson equation in the spherical coordinate chart, which does not include the origin, but in Cartesian coordinates, which cover the origin too, it acquires a source term. Similarly, the solution of the diffeomorphism constraint in spherical coordinates depends on the integration constant $A$, which corresponds, in Cartesian coordinates, to a Dirac-delta source term for the constraint. It is clear now how this ruins the predictivity of the theory: one is free to specify a source term like~(\ref{DiffeoConstraintWithDiracDeltaSource}) as a function of time, and no dynamical equation can fix it for us. One may be interested in this exercise, to model for example some collapsed  matter which has some expansion/contraction, but is concentrated in a small region that we want to approximate as pointlike. However, for the present problem of modelling the gravitational collapse of a distribution of matter, it is clear that we have to require that the effective value of the integration constants $A$ and $m$ at the poles is zero.

\newpage

\section{Appendix: Region of positivity of $\mathscr P$ and on-shell surfaces}
\label{MordorAppendix}

\subsection{$\mathscr P > 0$ region}

\newpage

%

\subsection{On-shell surfaces}\label{AppendixOnshellSurfaces}

\begin{figure}[ht!]
\includegraphics[width=\widfigs\textwidth]{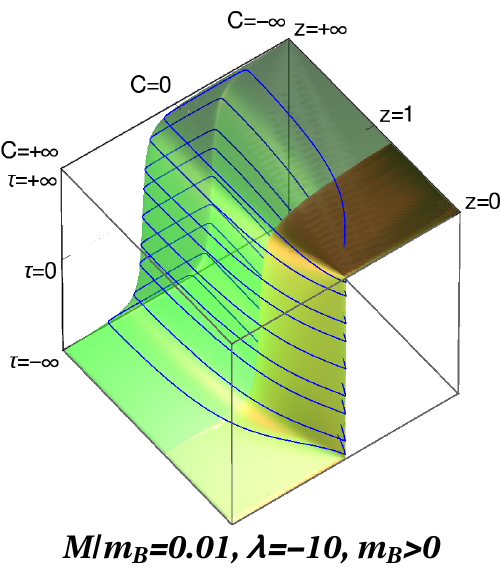}~~~~~~~~~~~~~~~~\includegraphics[width=\widfigs\textwidth]{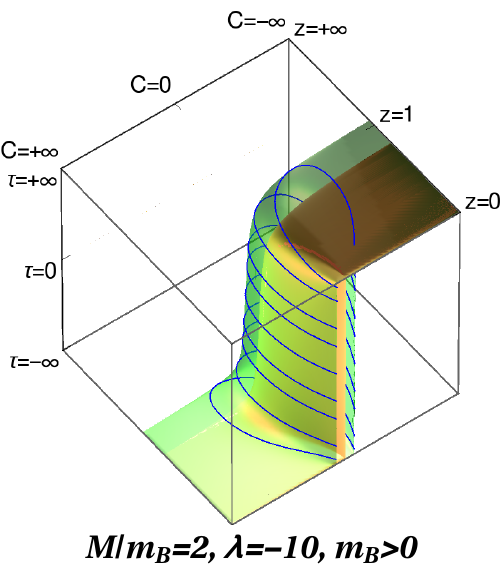}
\\
\includegraphics[width=\widfigs\textwidth]{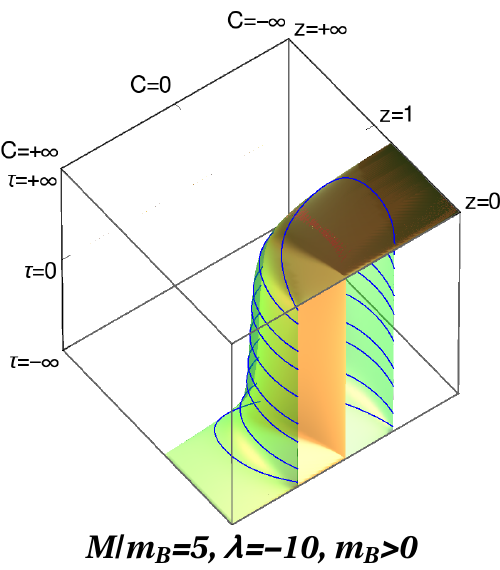}~~~~~~~~~~~~~~~~\includegraphics[width=\widfigs\textwidth]{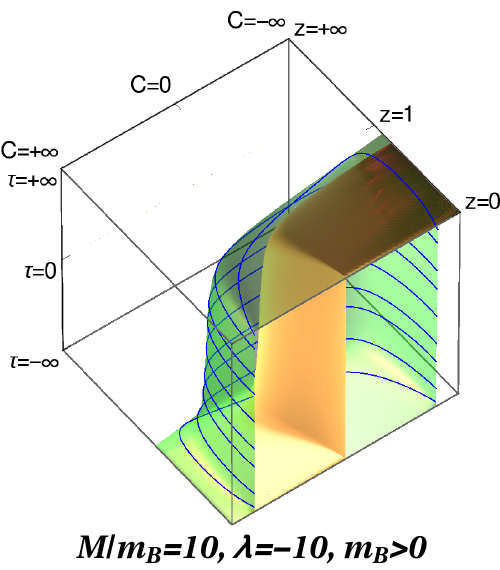}
\caption[On-shell surfaces for $\lambda<0$, $m_\st{B}>0$]{\footnotesize
The surface $\mathscr P (z_a) =0$  of Fig.~\ref{MordorFig1}
 for $m_\st{B}>0$, $\lambda = -10$ (in yellow), together with the on-shell surface (transparent green), for four choices of the ratio $M_a/m_\st{B}$.}
\label{Mordor_Onshell_Fig_1}
\end{figure}

\begin{figure}[ht!]
\includegraphics[width=\widfigs\textwidth]{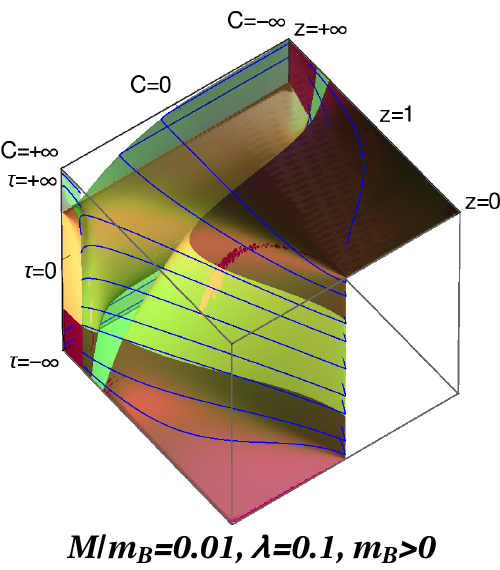}~~~~~~~~~~~~~~~~\includegraphics[width=\widfigs\textwidth]{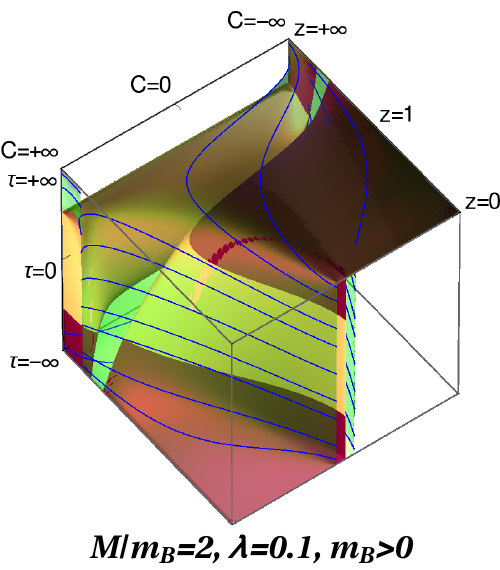}
\\
\includegraphics[width=\widfigs\textwidth]{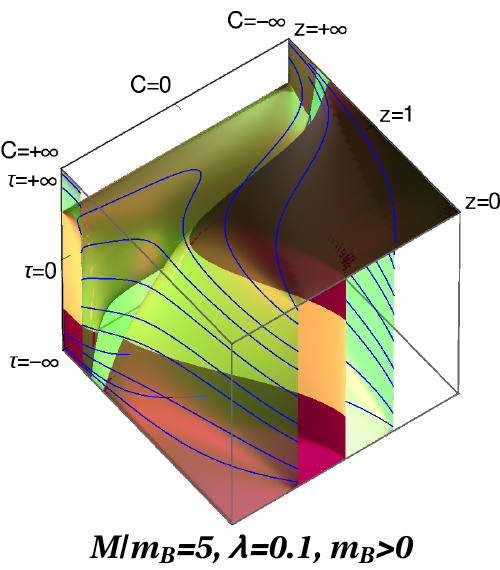}~~~~~~~~~~~~~~~~\includegraphics[width=\widfigs\textwidth]{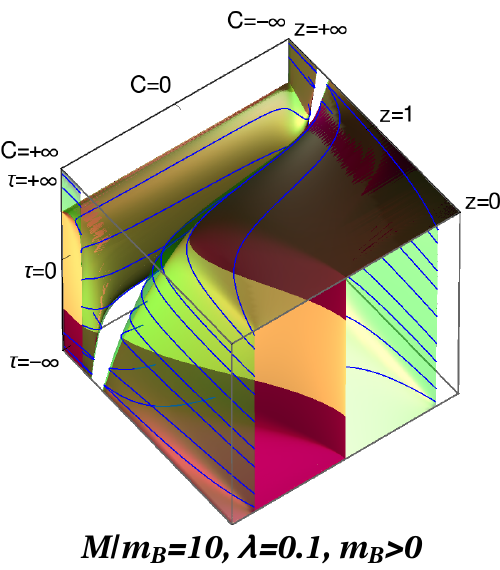}
\caption[On-shell surfaces for $\lambda>0$, $m_\st{B}>0$]{\footnotesize
The surface $\mathscr P (z_a) =0$  of Fig.~\ref{MordorFig2}
 for $m_\st{B}>0$, $\lambda = 0.1 >0$ (yellow/red) and the on-shell surface (in transparent green), for four choices of the ratio $M_a/m_\st{B}$.}
\label{Mordor_Onshell_Fig_2}
\end{figure}

\begin{figure}[ht!]
\begin{minipage}{\textwidth}\center
\includegraphics[width=\widfigs\textwidth]{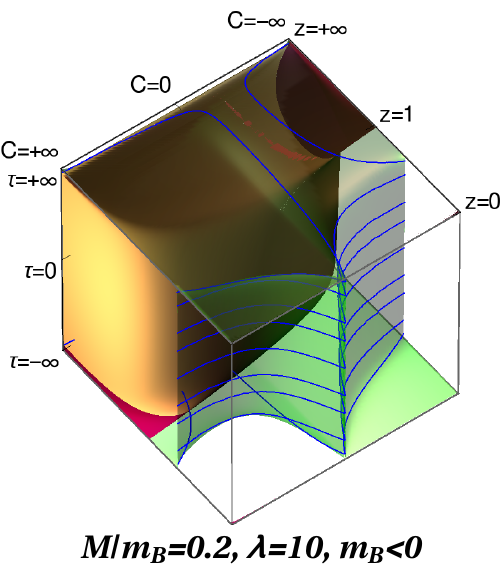}~~~~~~~~~~~~~~~~\includegraphics[width=\widfigs\textwidth]{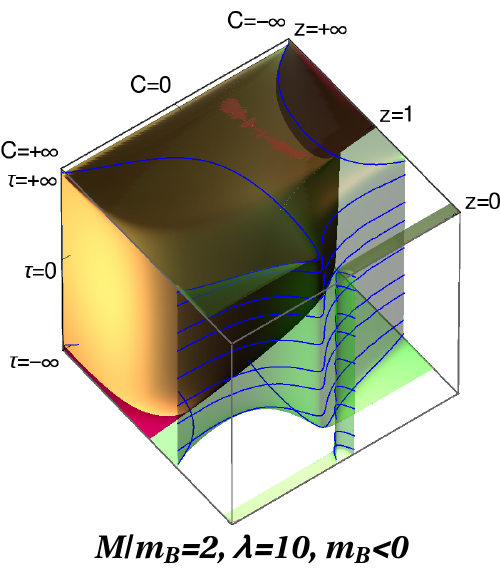}
\\
\includegraphics[width=\widfigs\textwidth]{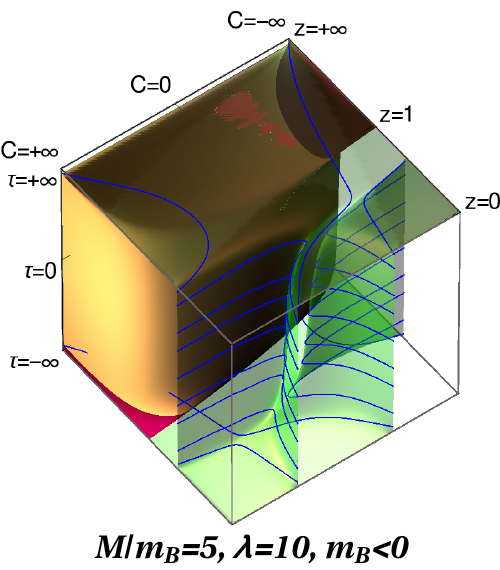}~~~~~~~~~~~~~~~~\includegraphics[width=\widfigs\textwidth]{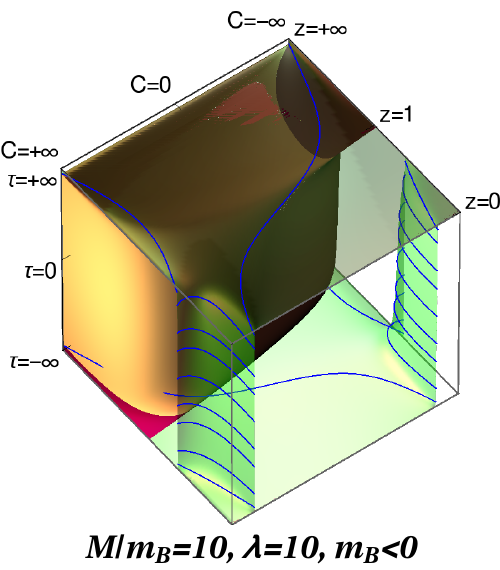}
\end{minipage}
\caption[On-shell surfaces for $\lambda>0$, $m_\st{B}<0$]{\footnotesize
The surface $\mathscr P (z_a) =0$  of Fig.~\ref{MordorFig3}
 for $m_\st{B}<0$, $\lambda = 10 >0$ (in yellow/red), together with the on-shell surface (in transparent green), for four choices of the ratio $M_a/m_\st{B}$.}
\label{Mordor_Onshell_Fig_3}
\end{figure}

\begin{figure}[ht!]
\begin{minipage}{\textwidth}\center
\includegraphics[width=\widfigs\textwidth]{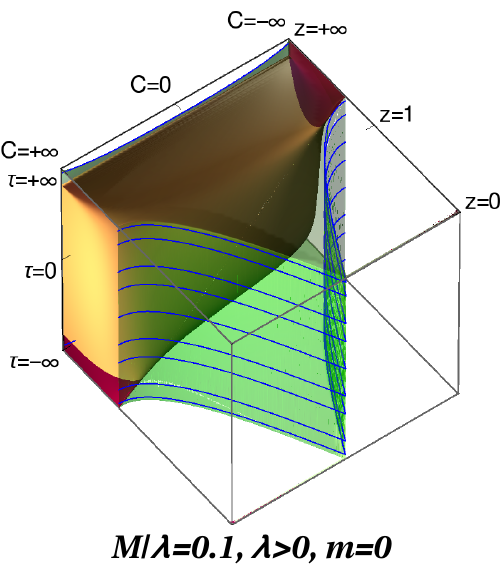}~~~~~~~~~~~~~~~~\includegraphics[width=\widfigs\textwidth]{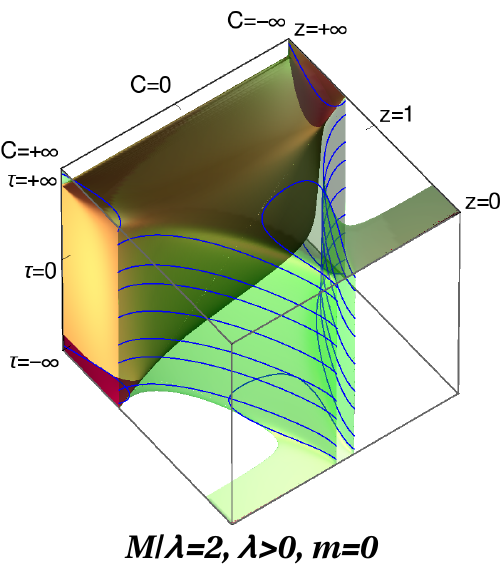}
\\
\includegraphics[width=\widfigs\textwidth]{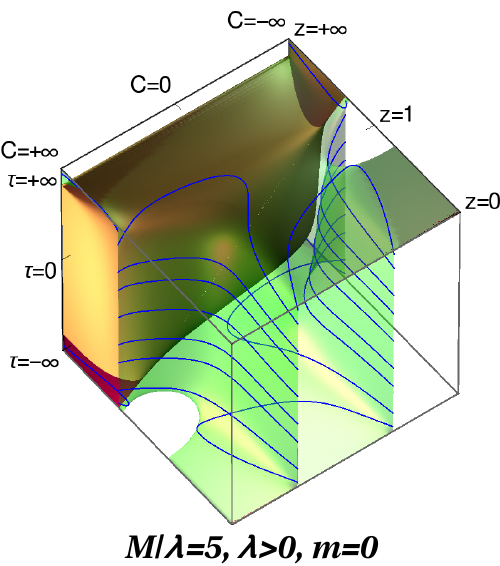}~~~~~~~~~~~~~~~~\includegraphics[width=\widfigs\textwidth]{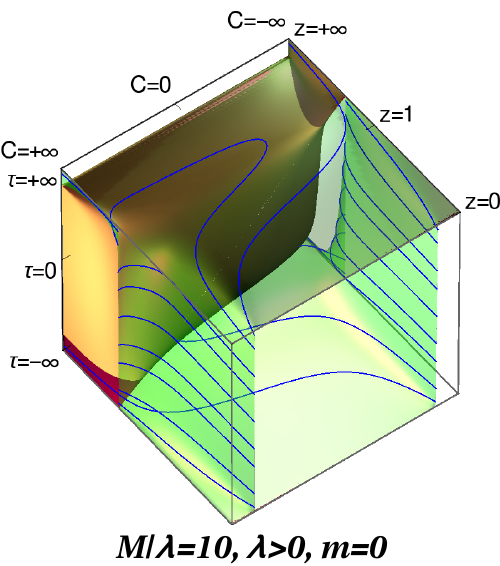}
\end{minipage}
\caption[On-shell surfaces for $\lambda>0$, $m_\st{B}=0$]{\footnotesize
The surface $\mathscr P (z_a) =0$  of Fig.~\ref{MordorFig4}
 for $m_\st{B}=0$ (in yellow/red), together with the on-shell surface (in transparent green), for four choices of the ratio $M_a/\lambda$.}
\label{Mordor_Onshell_Fig_4}
\end{figure}

\begin{figure}[ht!]
\begin{minipage}{\textwidth}\center
\includegraphics[width=\widfigs\textwidth]{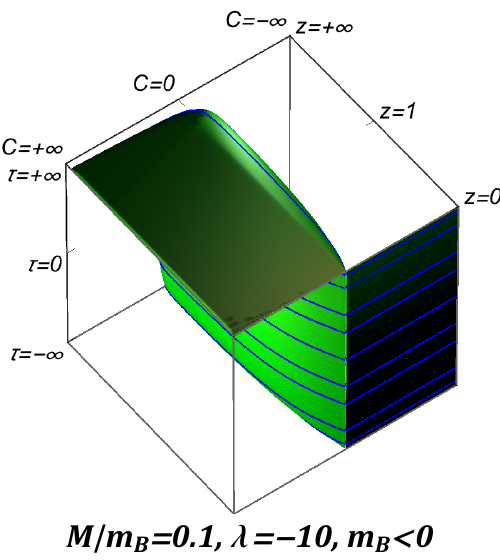}~~~~~~~~~~~~~~~~\includegraphics[width=\widfigs\textwidth]{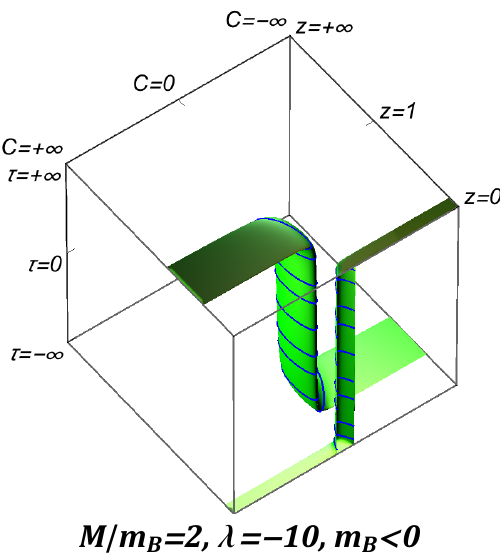}
\\
\includegraphics[width=\widfigs\textwidth]{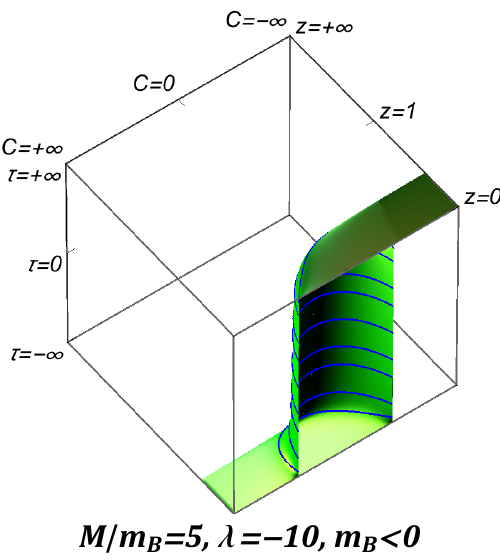}~~~~~~~~~~~~~~~~\includegraphics[width=\widfigs\textwidth]{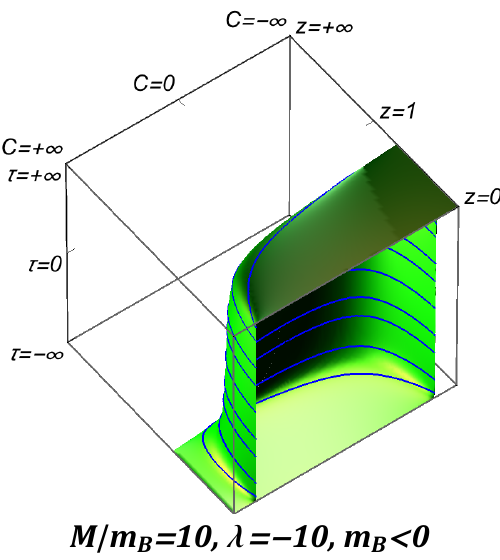}
\end{minipage}
\caption[On-shell surface  for $\lambda<0$, $m_\st{B}<0$]{\footnotesize
On-shell surface for negative $\lambda$ and $m_\st{B}$, for four choices of the ratio $M_a/m_\st{B}$. In this case there is no excluded region because for this choice of signs of $\lambda$ and $m_\st{B}$ all values of the parameters are admissible.}
\label{Mordor_Onshell_Fig_5}
\end{figure}

\begin{figure}[ht!]
\begin{minipage}{\textwidth}\center
\includegraphics[width=\widfigs\textwidth]{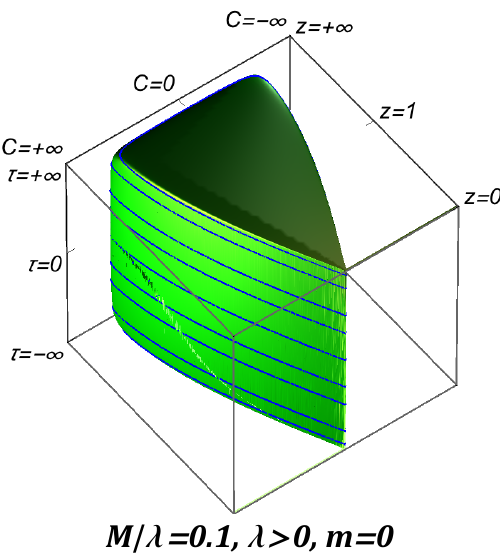}~~~~~~~~~~~~~~~~\includegraphics[width=\widfigs\textwidth]{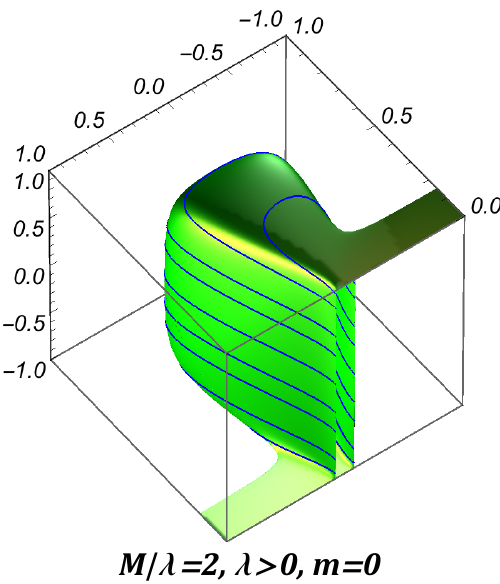}
\\
\includegraphics[width=\widfigs\textwidth]{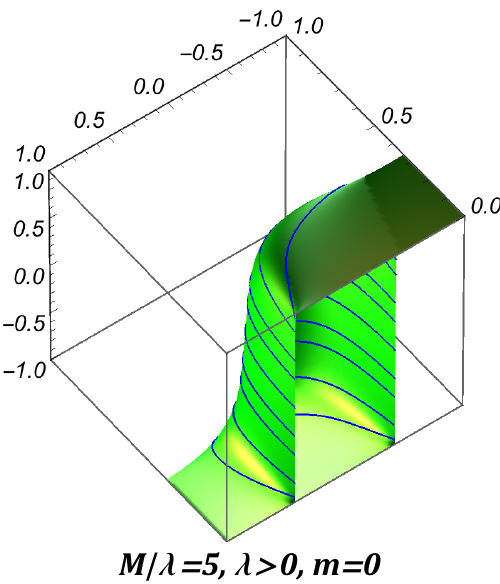}~~~~~~~~~~~~~~~~\includegraphics[width=\widfigs\textwidth]{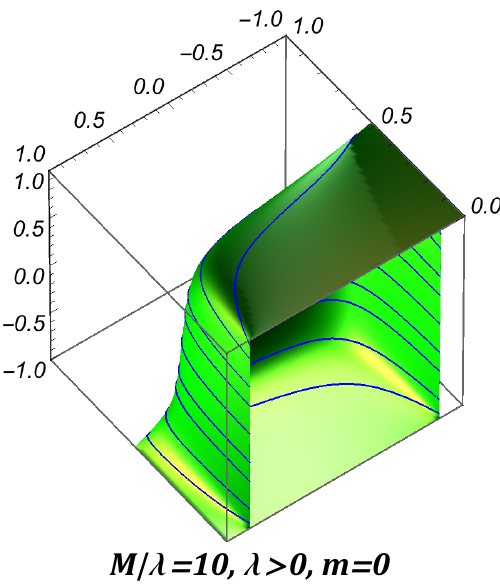}
\end{minipage}
\caption[On-shell surface  for $\lambda<0$, $m_\st{B}=0$]{\footnotesize
On-shell surface for negative $\lambda$ and $m_\st{B}=0$, for four choices of the ratio $M_a/|\lambda|$. In this case too there is no excluded region because with $m_\st{B}=0$ and for this choice of signs of $\lambda$ all values of the parameters are admissible.}
\label{Mordor_Onshell_Fig_6}
\end{figure}

\end{document}